\documentclass[a4paper,11pt,twoside,onecolumn,final,openright]{book}

\usepackage[doublespace,final]{thesis}

\usepackage{subfigure}
\usepackage{moreverb}
\usepackage{amssymb}
\usepackage[doublespace,final]{thesis}
 \usepackage{amsmath}
\usepackage [english]{babel}
\usepackage [autostyle, english = american]{csquotes}
\usepackage{listings}
\usepackage{color}
\usepackage{textcomp}
\usepackage{algpseudocode}
\usepackage{algorithm}
\usepackage{lmodern}
\usepackage{enumerate}

\definecolor{listinggray}{gray}{0.98}
\definecolor{lbcolor}{rgb}{0.98,0.98,0.98}
\newenvironment{changemargin}[2]{%
\begin{list}{}{%
\setlength{\leftmargin}{#1}%
\setlength{\rightmargin}{#2}%
}%
\item[]}
{\end{list}}

\begin{document}
\begin{changemargin}{-0.25cm}{-0.5cm}   

\def\date{September 2014}
\def\titulo{An Elastic Middleware Platform for Concurrent and Distributed Cloud and MapReduce Simulations}

\hypersetup{colorlinks,
   debug=false,
   linkcolor=blue,  
   citecolor=red,  
   urlcolor=blue,   
   bookmarksopen=true,
   pdftitle={\titulo},
   pdfauthor={Pradeeban Kathiravelu},
   pdfsubject={Cloud2Sim},
   pdfkeywords={Master Thesis}
}

  %
  %

\thispagestyle{empty}

\begin{singlespace}
\vbox to\textheight{%
\vskip-1.3in
\vbox to10mm{\LARGE\sl
\vfil}

\hskip-5mm\vbox to50mm{
\vfil

\begin{tabular}{l}
\includegraphics[width=4.5cm]{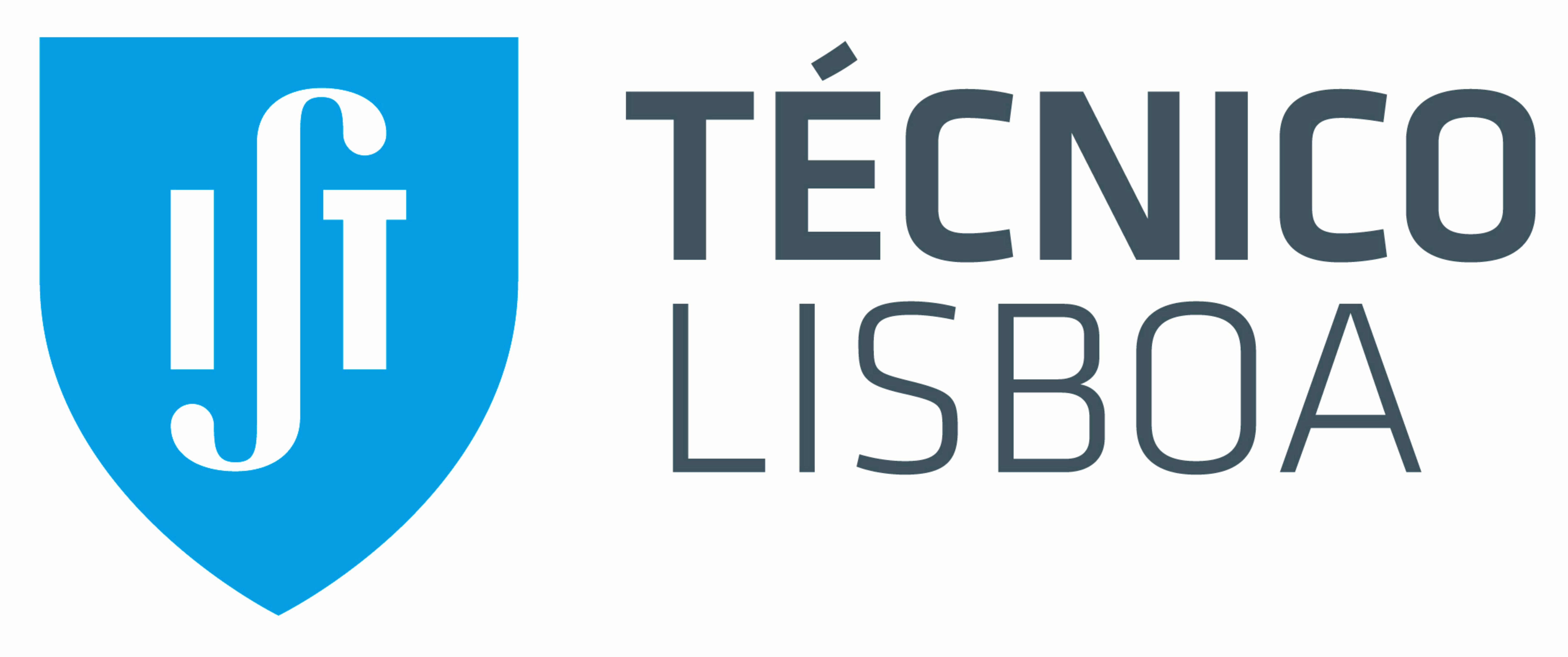}
\end{tabular}
\vfil
\vfil
}%
\vskip13mm
\vbox to15mm{\LARGE\sl
\vfil
\vfil
}%
\vskip6mm
\vbox to25mm{\LARGE\bf
\vfil
\begin{center}
\titulo
\end{center}
\vfil
}%
\vskip10mm
\vbox to25mm{\large
\vfil
\begin{center}
{\Large\bf Pradeeban Kathiravelu}\\   
\end{center}
\vfil
}%
\vskip8mm
\vbox to8mm{\large
\vfil
\centerline{Thesis to obtain the Master of Science Degree in}
\vskip2mm
\centerline{{\Large\bf Information Systems and Computer Engineering} }
\vfil
}%
\vskip10mm

\begin{center}
\begin{tabular}{p{0.2\textwidth}l}
Supervisor: & Doctor Luís Manuel Antunes Veiga\\
\end{tabular}
\end{center}

\vbox to8mm{\large
\vfil

\begin{center}
{\Large\bf Examination Committee}\\
\end{center}
\vfil
}%

\vbox to3mm{\large
\vfil
\begin{center}
\begin{tabular}{p{0.3\textwidth}l}
Chairperson: & Doctor José Carlos Alves Pereira Monteiro\\
Supervisor: & Doctor Luís Manuel Antunes Veiga\\
Member of the Committee: & Doctor Ricardo Jorge Freire Dias\\
\end{tabular}
\end{center}
\vfil
}%
\vskip23mm
\vbox to4mm{\Large\bf
\vfil
\begin{center}
\date
\end{center}
\vfil
}%
}
\end{singlespace}
\newpage

  %
  %

\chapter*{Acknowledgements}
\thispagestyle{empty}


$Cloud^{2}Sim$ was a brain-child of Prof. Luís Manuel Antunes Veiga who proposed the topic and accepted my interest to work on this topic. I would like to thank my supervisor for leading me throughout the project with his creative suggestions on making the thesis better and guidance throughout the thesis. His lectures and guidance were always motivating.

I would like to thank Prof. João Coelho Garcia for his motivations and inspirations. Prof. Luis Rodrigues helped us at various occasions. I would also like to thank Prof. Johan Montelius for his leadership and guidance, during my third semester at KTH Royal Institute of Technology. I would like to thank all the professors who taught us at IST and KTH Royal Institute of Technology during my master studies.

Erasmus Mundus offered us an opportunity to come from different countries and study at these wonderful institutes. I would like to extend my thanks to the Education, Audiovisual and Culture Executive Agency (EACEA) of the European Union, and EMDC consortium for selecting me for the scholarship. My thanks goes to all my friends who shared the same journey with me with Erasmus Mundus, making my studies a pleasant experience.

This work was partially supported by Erasmus Mundus Master Courses Scholarship (Category A), provided by EACEA.

\vfill
\begin{flushright}
  \begin{minipage}{8cm}
    \begin{center}
      Lisboa, \today

      Pradeeban Kathiravelu
    \end{center}
  \end{minipage}
\end{flushright}

\cleardoublepage

\chapter*{European Master in Distributed Computing (EMDC)}
This thesis is a part of the curricula of the European Master in Distributed Computing, a cooperation between KTH Royal Institute of Technology in Sweden, Instituto Superior Técnico (IST) in Portugal and Universitat Politècnica de Catalunya (UPC) in Spain. This double degree master program is supported by the Education, Audiovisual and Culture Executive Agency (EACEA) of the European Union.

My study track during the master studies of the two years is as follows: \\
First year: Instituto Superior Técnico, Universidade de Lisboa \\
Third semester: KTH Royal Institute of Technology \\
Fourth semester (Thesis): INESC-ID/Instituto Superior Técnico, Universidade de Lisboa

\cleardoublepage
  %
  %

\chapter*{}
\thispagestyle{empty}

\vfill
\mbox{}
\vfill\Large
\begin{flushright}
  \begin{minipage}{8cm}
    \begin{center}

To my parents and my teachers

    \end{center}
  \end{minipage}
\end{flushright}
\normalsize\vfill

\cleardoublepage

  %
  %

\chapter*{Resumo}
\thispagestyle{empty}
A investigação no contexto da Computação em Nuvem envolve um grande número de entidades, como utilizadores, clientes, aplicações, e máquinas virtuais. Devido ao acesso limitado e disponibilidade variável de recursos, os investigadores testam os seus protótipos em ambientes de simulação, em vez dos ambientes reais na Nuvem. Contudo, ambientes de simulação nuvem actuais, como CloudSim e EmuSim são executados sequencialmente. Um ambiente de simulação mais avançado poderia ser criado estendendo-os, aproveitando as mais recentes tecnologias, bem como a disponibilidade de computadores multi-core e os clusters nos laboratórios de investigação. Embora a computação tenha evoluído com a programação multi-core, o paradigma MapReduce e as plataformas de middleware, as simulações de escalonamento e gestão de recursos na nuvem e de MapReduce ainda não exploram estes avanços. Neste trabalho, desenvolvemos o Cloud2Sim, atacando esta falta de correspondência entre simulações e tecnologia atual que elas tentam simular.

Propomos um simulador de nuvem, Cloud2Sim, concorrente e distribuído, estendendo o simulador CloudSim, usando o armazenamento chave-valor em memória distribuída Hazelcast. Fornecemos uma avaliação das implementações de MapReduce no  Hazelcast e Infinispan, distribuindo de forma adaptativa a execução de um cluster, fornecendo também meios para simulação de execuções MapReduce. A nossa solução dinâmica escala as simulações de nuvens e MapReduce para vários nós que executam Hazelcast e Infinispan, com base na carga. O modelo de execução distribuído e a solução de escalonamento adaptativo podem tornar-se um middleware geral para auto-scaling numa infrastrutura multi-cliente (multi-tenanted).

\newpage

  %
  %

\chapter*{Abstract}
\thispagestyle{empty}

Cloud Computing researches involve a tremendous amount of entities such as users, applications, and virtual machines. Due to the limited access and often variable availability of such resources, researchers have their prototypes tested against the simulation environments, opposed to the real cloud environments. Existing cloud simulation environments such as CloudSim and EmuSim are executed sequentially, where a more advanced cloud simulation tool could be created extending them, leveraging the latest technologies as well as the availability of multi-core computers and the clusters in the research laboratories. While computing has been evolving with multi-core programming, MapReduce paradigms, and middleware platforms, cloud and MapReduce simulations still fail to exploit these developments themselves. This research develops $Cloud^{2}Sim$, which tries to fill the gap between the simulations and the actual technology that they are trying to simulate.

First, $Cloud^{2}Sim$ provides a concurrent and distributed cloud simulator, by extending CloudSim cloud simulator, using Hazelcast in-memory key-value store. Then, it also provides a quick assessment to MapReduce implementations of Hazelcast and Infinispan, adaptively distributing the execution to a cluster, providing means of simulating MapReduce executions. The dynamic scaler solution scales out the cloud and MapReduce simulations to multiple nodes running Hazelcast and Infinispan, based on load. The distributed execution model and adaptive scaling solution could be leveraged as a general purpose auto scaler middleware for a multi-tenanted deployment.

\newpage

  %
  %

\chapter*{Palavras Chave \\ Keywords}
\thispagestyle{empty}

\section*{Palavras Chave}
{\large 

\noindent Computação em Nuvem

\noindent Simulação

\noindent \emph{Auto Scaling}

\noindent \emph{MapReduce}

\noindent Computação Voluntária

\noindent Partilha de Ciclos

\noindent Execução Distribuída

}

\section*{Keywords}

{\large 

\noindent Cloud Computing

\noindent Simulation

\noindent Auto Scaling

\noindent MapReduce

\noindent Volunteer Computing

\noindent Cycle Sharing

\noindent Distributed Execution

}

\vfill

\cleardoublepage

  %
  %

\pagestyle{plain}
\pagenumbering{roman}

  %
  %


\def\contentsname{Index}
\tableofcontents
\newpage

\listoffigures
\newpage

\listoftables

\chapter*{Abbreviations}
\noindent {\em ASP} --- Adaptive Scaler Probe

\noindent {\em AWS} --- Amazon Web Services

\noindent {\em CFD} --- Computational Fluid Dynamics

\noindent {\em EC2} --- Elastic Computing Cloud

\noindent {\em FCFS} --- First-Come First-Served

\noindent {\em GC} --- Garbage Collection

\noindent {\em GPU} --- Graphics Processing Unit

\noindent {\em GT} --- Globus Toolkit

\noindent {\em HTC} --- High Throughput Computing

\noindent {\em IaaS} --- Infrastructure-as-a-Service

\noindent {\em IAS} --- Intelligent Adaptive Scaler

\noindent {\em IDC} --- Internet Data Center

\noindent {\em IMDG} --- In-Memory Data Grid

\noindent {\em JVM} --- Java Virtual Machine

\noindent {\em MI} --- Million Instructions

\noindent {\em MIPS} --- Million Instructions Per Second

\noindent {\em MM} --- Matchmaking

\noindent {\em MVCC} --- Multiversion Concurrency Control

\noindent {\em PaaS} --- Platform-as-a-Service

\noindent {\em P2P} --- Peer-to-Peer

\noindent {\em RR} --- Round Robin

\noindent {\em TCL} --- Tool Command Language

\noindent {\em VM} --- Virtual Machine

\noindent {\em VMM} --- Virtual Machine Manager

\noindent {\em VO} --- Virtual Organization

\vfill

\cleardoublepage

  %
  %

\pagestyle{plain}
\pagenumbering{roman}

\cleardoublepage

  %
  %

\addtolength{\textwidth}{4mm}
\addtolength{\textheight}{4mm}

\pagenumbering{arabic}
\pagestyle{headings}
\chapter{Introduction}
Simulations empower the researchers with an effective and quicker way to test the prototype developments of their research. As cloud computing~\cite{rimal2009taxonomy} environments consist of data centers and applications distributed on a planetary-scale~\cite{agarwal2010volley}, cloud simulations are used in evaluating architectures, algorithms, topologies, and strategies that are under research and development, tackling many issues such as resource management, application scheduling, load balancing, workload execution, and optimizing energy consumption. While the exact environment of the cloud platform may not be accessible to the developers at the early stages of development, simulations~\cite{cloudsim,peersim,oversim} give an overall idea on the related parameters, resource requirements, performance, and output. With the increasing complexity of the systems that are simulated, cloud simulations are getting larger and the larger simulations tend to take longer and longer time to complete being run in a single node.

Java in-memory data grids provide a distributed execution and storage model for problems in the grid-scale. They offer scalability and seamless integration with persistent storage. Hazelcast~\cite{hazelcast}, Infinispan~\cite{infinispan}, Terracotta BigMemory\footnote{\url{http://terracotta.org/products/bigmemory}}, and Oracle Coherence~\cite{seovic2010oracle} are some of the currently most used platforms for distributed execution and storage~\cite{ferrante2010java,el2012grid,mohanty2013extracting}. Using these platforms, users can create data grids and distributed cache, on the utility computers, to execute much larger jobs that cannot be run on any single computer, or that would take a huge time to execute often with a slower response.

Exploiting the existing simulation approaches that are heavily centralized, and the distributed execution platforms, cloud simulations can be made distributed, such that they can be able to utilize the computer clusters in the research labs. Distributed simulations can enable larger simulations to execute in a shorter time with a better response, whilst making it possible to simulate scenarios that may not even be possible on a single instance.

\section{Problem Statement}
Cloud simulation environments require a considerable amount of memory and processing power to simulate a complex cloud scenario. Processors are increasingly becoming more powerful with multi-core architectures and the computing clusters in the research laboratories themselves can be used to run complicated large simulations in a distributed manner, as in BOINC derivatives~\cite{silva2008nuboinc}. However, current simulation tools provide very limited support to utilize these resources, as they are mostly written with a sequential execution model targeting to run on a single server.

Utilizing distributed computers to share the cycles to the simulation, as required by the simulation, would enable simulating bigger and more complex scenarios that cannot be simulated effectively in a single node, or it could be a very time consuming execution. While cycle sharing and volunteer computing are used in scientific research and grid computing projects, these models are not widely utilized to provide computing resources for cloud simulations. Moreover, when the resource providers are inside a trusted private network such as a research lab, security concerns related to cycle sharing can be considered lightly. Hence, the cycle sharing model can be leveraged to operate in a private cluster to provide a scalable middleware platform for simulations.

Cloud simulations are becoming resource-hungry and time-consuming, as the cloud systems they attempt to simulate are getting more and more complicated. Cloud simulators simulate the systems involving data centers, hosts, virtual machines, and cloudlets, the applications that run top of virtual machines. The cloudlets often execute independent workloads that do not depend on the workloads of the other cloudlets. Hence, a cloud simulator can be made distributed, where the executions such as the cloudlet workloads and the scheduling components that schedule the workloads to the virtual machines and hosts, can be made to execute in a distributed manner. Cloud simulators should be able to execute effectively in a single node as well as from multiple distributed instances. 

In-memory data grids enable an effective distribution of storage and execution, which attempt to minimize the possible overheads from a distributed execution, while giving an illusion of a single large computer over multiple computer nodes. Distribution approaches such as a client-server architecture will not provide a seamless integration with the system to function as a single node as well as in a distributed environment, due to their architecture and communication delays. Hence, functionality and scalability of the cloud simulators can be extended using the in-memory data grids, while maintaining the accuracy and performance of the simulator. 

Existing cloud simulators also lack the ability to simulate MapReduce tasks, while there are simulators just specific to MapReduce. However, a MapReduce simulator can be implemented along with the cloud simulator, to simulate complex scenarios involving MapReduce tasks and cloud applications such as load balancing the MapReduce tasks into different data centers~\cite{wang2013g} and power-aware resource scheduling~\cite{zhou2013carbon}. The result of this research led to the development of $Cloud^{2}Sim$, a concurrent and distributed cloud simulator, with extended functionality including the ability to dynamically scale with the load of the simulations, and simulate MapReduce applications executing on clouds.

\section{Thesis Objectives and Contributions}
This project researches and implements a concurrent and distributed cloud simulator, named ``$Cloud^{2}Sim$'', by leveraging CloudSim~\cite{cloudsim,cloudgridsim} as the core simulation module, whilst taking advantage of the distributed shared memory provided by Hazelcast and in-memory key-value data grid of Infinispan.

The major contributions of this thesis are further detailed below:
\begin{itemize}
\item Concurrent and distributed architecture for Cloud and MapReduce Simulations.
\begin{itemize}
\item Design and implementation of a concurrent and distributed cloud simulator.
\item Design and implementation of a scalable and elastic middleware platform for the simulations.
\item Partitioning and scaling out of the cloud simulations and MapReduce executions to multiple nodes.
\item Implementations and evaluations of MapReduce simulations, based on Hazelcast and Infinispan MapReduce implementations.
\end{itemize}
\end{itemize}
\begin{itemize}
\item A generic adaptive scaling algorithm able to add nodes to the execution, in a cycle sharing fashion, to handle heavy-duty executions effectively.
\begin{itemize}
\item An adaptive scaler implementation, $IntelligentAdaptiveScaler$, for cloud and MapReduce simulations with elasticity and fail-over.
\item A detailed model and study on scalability and performance enhancements by a distributed execution.
\end{itemize}
\end{itemize}
\begin{itemize}
\item Support for multi-tenanted cloud and MapReduce simulations with a $Coordinator$ design and implementation.
\begin{itemize}
\item Configurable CloudSim simulations.
\item Ability to scale out simulations to real elastic cloud environments.
\end{itemize}
\end{itemize}

\section{Publications}
The work described in this paper has been partially described (with some preliminary results) in the below technical publications.
\begin{itemize}
\item \emph{IEEE 22nd International Symposium on Modeling, Analysis and Simulation of Computer and Telecommunication Systems (MASCOTS 2014)}. The paper is: Pradeeban Kathiravelu and Luis Veiga, \emph{``Concurrent and Distributed CloudSim Simulations''}~\cite{MASCOTS}.
\item \emph{IEEE/ACM 7th International Conference on Utility and Cloud Computing (UCC 2014)}. The paper is: Pradeeban Kathiravelu and Luis Veiga, \emph{``An Adaptive Distributed Simulator for Cloud and MapReduce Algorithms and Architectures''}~\cite{UCC}.
\end{itemize}

\section{Organization of the Thesis}
In the upcoming chapters in this dissertation, we will further analyze the architecture and design of $Cloud^{2}Sim$ and how it is designed and implemented leveraging CloudSim as the core simulation module. Chapter~\ref{chap:rw} discusses the cloud and MapReduce simulators, discussing the architectural details of CloudSim in detail, and continues to discuss the in-memory data grids (IMDG) and cycle-sharing architectures and systems in the latter part of the chapter.

Chapter~\ref{chap:sa} discusses the solution architecture of $Cloud^{2}Sim$, the design, and how CloudSim and in-memory data grids are optimized and leveraged to design an elastic middleware platform for concurrent and distributed cloud and MapReduce simulations. Chapter~\ref{chap:imp} discusses the implementation of the concurrent and distributed cloud and MapReduce simulator and looks how the execution is distributed in detail. It further addresses the scalability of the implementation, and how the scalable middleware platform is implemented, discussing the dynamic scaling of the implementation.

$Cloud^{2}Sim$ was benchmarked against CloudSim and $Cloud^{2}Sim$ was evaluated on multiple nodes. The evaluation results are discussed in Chapter~\ref{chap:eval}, discussing the CloudSim and MapReduce simulation experiments carried out on $Cloud^{2}Sim$, along with some comparison of $Cloud^{2}Sim$ with related projects. Finally, Chapter~\ref{chap:cf} will drive us to the conclusion of this research discussing its current state and the possible future enhancements.

\chapter{Related Work}
\label{chap:rw}

We will discuss the related research, projects, and technology in this chapter. Considering the distributed cloud simulation as a major contribution of this research, Section~\ref{sec:2cloud} will discuss the cloud simulators in detail. Further, we will go through the MapReduce simulators in Section~\ref{sec:2ms}, as our work also contains a distributed execution environment for MapReduce implementations and simulations. Distributed execution environments such as in-memory data grids and distributed caches are proposed as the enabling means for the distribution. Hence, we will also go through the related distributed in-memory key-value stores and data grids in Section~\ref{sec:2de}. Moreover, the project proposes a cycle sharing based approach for the dynamic scaling, using the computers of the research labs, to schedule the execution of simulation activity tasks. Thus, in Section~\ref{sec:2cs}, the final section of the chapter, we will go through the cycle sharing solutions to conclude the study of related works.

  %
  %

\section{Cloud Simulators}
\label{sec:2cloud}
While some of the simulators are general-purpose, others focus on a narrower domain for simulation. CloudSim~\cite{cloudsim,buyya2009modeling,cloudgridsim}, EmuSim~\cite{emusim}, and GreenCloud~\cite{greencloud} are some of the mostly used general-purpose cloud simulation environments. MDCSim~\cite{lim2009mdcsim} and DCSim~\cite{dcsim} are simulators designed specifically for data center simulation. OverSim~\cite{oversim}, PlanetSim~\cite{garcia2005planetsim}, P2PSim~\cite{gil2003p2psim}, Overlay weaver~\cite{shudo2008overlay}, and PeerSim~\cite{peersim} are simulation toolkits for peer-to-peer and overlay networks.

Many grid simulators such as SimGrid~\cite{simgrid,legrand2003scheduling,casanova2008simgrid} evolved into cloud simulators, or have been extended into a cloud simulator. Grid computing consists of virtual organizations (VO) which are service provider entities sharing and following a set of access and management rules~\cite{foster2001anatomy,coppola2008virtual}. As both cloud and grid computing focus on the virtualization and resource allocation, the problems addressed by the simulators are of the similar nature. However, their use cases differ, as grid focuses more on huge batch tasks where clouds focus more on multiple smaller and online tasks with multi-tenancy~\cite{dillon2010cloud}. Hence, grid simulators require extensions to function as cloud simulators. Originally developed as GridSim, a grid simulation tool, CloudSim was later extended as a Cloud Simulation environment. GangSim~\cite{dumitrescu2005gangsim}, ChicSim~\cite{ranganathan2002decoupling}, MicroGrid~\cite{song2000microgrid}, and OptorSim~\cite{bell2003optorsim} are some other grid simulators.

We will address in greater detail the most relevant simulators such as CloudSim and SimGrid and analyze them comparatively in the end of the section.

\subsection{CloudSim}
Initially having GridSim as a major building block~\cite{cloudgridsim}, CloudSim was further developed by the CLOUDS laboratory as a cloud simulator on its own. CloudSim is frequently used by researchers, because of its extensibility and portability. Due to its modular architecture which facilitates customizations, it has been extended into different simulation tools such as CloudAnalyst~\cite{cloudanalyst}, WorkflowSim\cite{chen2012workflowsim}, and NetworkCloudSim~\cite{ncloudsim}.

CloudSim defines the parameters of the cloud environments such as hosts, VMs, applications, and data centers by the instances of different classes. $Datacenter$ is the resource provider which simulates infrastructure-as-a-service. Multiple hosts are created inside data centers~\cite{buyya2009modeling}. There should be at least one data center in the system for CloudSim to start execution. $DatacenterBroker$ is responsible for application scheduling and coordinating the resources. $DatacenterBroker$ functions as the coordinating entity of resources and user applications. A single broker or a hierarchy of brokers can be initiated depending on the simulation scenario. 

Figure~\ref{fig:scheduling} shows how a cloud environment is represented by the architecture of CloudSim in a high level, focusing the resource scheduling. CPU unit is defined by $Pe$ (Processing Element) in terms of millions of instructions per second (MIPS). Multi-core processors are created by adding multiple $Pe$ objects to the list of Processing Elements. All processing elements of the same machine have the same processing power (MIPS). Processing elements are the shared resources and cloudlets represent the applications that share these resources among them. Status of a processing element can be FREE (1), BUSY/Allocated (2), or FAILED (3) indicating its availability for the cloudlet.

\begin{figure}[!htbp]
\begin{center}
 \resizebox{0.6\columnwidth}{!}{
  \includegraphics[width=0.6\textwidth]{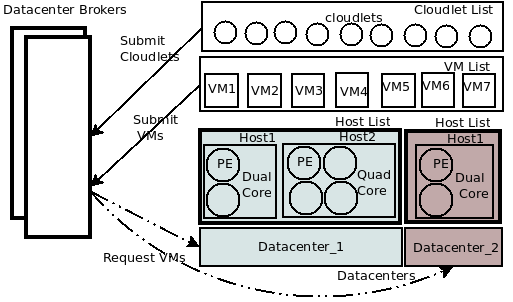}
 }
\end{center}
 \caption{CloudSim scheduling operations}
 \label{fig:scheduling}
\end{figure}

Each of the VMs is assigned to a host. Each cloudlet is assigned to a VM, and the processing elements are shared among the VMs in a host and among the executing cloudlets in the VMs. Complicated real-world cloud scenarios can be simulated by appropriately extending the broker or the other classes. Virtual machines and cloudlets are created and added to the respective lists. Once the simulation is started, the list of cloudlets and virtual machines are submitted to the broker. The broker handles the allocation of VMs to the hosts and cloudlets to the VMs, and leads and drives the simulation behavior such as deciding which of the available cloudlets to be executed next.

CloudSim can be used to model a dynamic scenario where elements are added and removed at run time. Simulations can be paused and resumed as well. CloudSim can model energy-aware data centers, that can optimize the power consumptions of the data centers~\cite{beloglazov2012optimal}. Support for user defined policies make CloudSim customizable and extensible for users' requirements. Extensions to CloudSim tend to address the limitations of CloudSim, or add more features to it. NetworkCloudSim enables modeling parallel applications such as MPI and workflows in CloudSim~\cite{ncloudsim}. WorkflowSim simulates scientific workflows, through a higher level workflow management layer~\cite{chen2012workflowsim}.

\subsection{SimGrid}
SimGrid~\cite{simgrid} is a toolkit initially developed for simulation of application scheduling. As a generic versatile simulator for large scale distributed computing, SimGrid offers four APIs - two APIs for researchers, and two more APIs for developers~\cite{casanova2008simgrid}. Figure~\ref{fig:simgrid} elaborates the SimGrid APIs and architecture. SimDag API allows the researchers to simulate scheduling heuristics. MSG interface lets the researchers analyze concurrent sequential processes (CSP), and rapidly prototype distributed algorithms~\cite{legrand2006simgrid}. GRAS (Grid Reality And Simulation) is a user API, that facilitates development of distributed applications. SMPI is the other user API, that lets users simulate MPI applications and heterogeneous platforms~\cite{legrand2006simgrid}. SimGrid approximates the behavior of the TCP networks, using a flow-level approach~\cite{fujiwara2007speed}. SimGrid has a higher scalability than most of the grid simulators, including all the general-purpose grid simulators referred in this thesis, such as ChicSim, MicroGrid, GridSim, and OptorSim~\cite{casanova2008simgrid}. 

\begin{figure}[!htbp]
\begin{center}
 \resizebox{0.5\columnwidth}{!}{
  \includegraphics[width=0.5\textwidth]{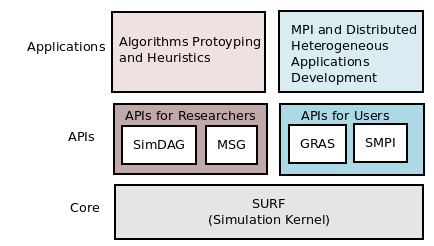}
 }
\end{center}
 \caption{SimGrid Architecture}
 \label{fig:simgrid}
\end{figure}

Simulation environments have a trade-off of accuracy/speed~\cite{velho2009accuracy}, producing faster less-accurate simulators and slower accurate simulators. Further researches focus on enhancing the speed, accuracy, and scalability of the existing simulators. Using dynamic routing, a research attempts to increase the scalability of SimGrid further, to simulate more nodes~\cite{de2009improving}. An extension has exploited SimGrid as a fast, scalable, and effective simulator for volunteer computing~\cite{donassolo2010fast}.

\subsection{Discrete Event Simulation Libraries}
Simulation of discrete events has been one of the early research efforts. Although a complete implementation was lacking, many parallel and distributed discrete event simulators were proposed and researched in late 1980's~\cite{misra1986distributed,fujimoto1990parallel}. Most of these early researches proposed theorems, algorithms, and strategies for an effective implementation for a parallel and distributed discrete event simulator, opposed to the sequential event simulators that were widely used by that time~\cite{misra1986distributed,fujimoto1990parallel,ferscha1998parallel,fujimoto1993parallel}. Further researches attempted to improve the proposed distributed algorithms~\cite{su1988variants}. Lately, discrete event simulation models are exploited in implementing some cloud and grid simulation tools. OMNeT++~\cite{varga2001omnet++,varga2008overview} based simulators and HyperSim~\cite{phatanapherom2003dynamic} are some examples of such simulators.

\paragraph*{iCanCloud:}
iCanCloud is a graphical cloud simulation platform, written in C++~\cite{nunez2012icancloud}. It is built on top of OMNeT++, which is a discrete event simulation library and graphical framework primarily used for building network simulators~\cite{varga2001omnet++}. Its major target is to facilitate high memory consuming applications such as those that require more than 2 GB of memory. It focuses on cost/performance trade-off, while letting the users simulate Amazon public clouds~\cite{nunez2012icancloud}. iCanCloud also promises distribution of simulation execution to multiple computers, based on MPI, as a work-in-progress.

\paragraph*{HyperSim:}
HyperSim is a simulation library for grid scheduling, claimed to be more scalable and faster than SimGrid and GridSim~\cite{phatanapherom2003dynamic}. Event graphs graphically represent the discrete event simulation models~\cite{buss1996modeling}. As a graphical simulator developed in C++, HyperSim requires the users to develop the respective event graphs, in order to construct any model on HyperSim~\cite{phatanapherom2003dynamic}.

\paragraph*{Multi-Agent Situated System (MMASS)-Based Simulations:}
Muli-agent systems consist of multiple agents, which are virtual or physical entities that can perceive their environment and act according to the defined logic~\cite{ferber1999multi}. Distributed artificial intelligence is a primary use case of multi-agent systems~\cite{ferber1999multi}. Multi-agent systems (MAS) are often represented by models. MMASS is such a model for MAS where the structures are explicitly defined~\cite{bandini2002dealing,bandini2006toward}. Distributed agent simulations are modeled and developed based on MMASS~\cite{bandini2006toward}.

Simulating a cloud is more intuitive than simulating a simple event based system. The initial research works carried on discrete event simulation libraries are not directly applicable to the current context of Cloud or MapReduce simulators, and are often specifications without implementation. There is no concurrent and distributed simulator implementation available as of now, which provides the same interface as CloudSim or other cloud simulators.

\subsection{Simulation of Energy-Aware Systems}
Design and simulation of energy-aware systems is a major requirement in many interdisciplinary fields, including electrical, communication, and systems engineering~\cite{shah2002energy,younis2002energy,beloglazov2012energy}. Optimizing the energy consumption is a major focus in cloud infrastructure, since power consumed by cloud data centers is enormous. Simulating energy-aware solutions has become part of the cloud simulators such as CloudSim~\cite{beloglazov2012energy}, as energy-aware simulation is becoming a major research interest for cloud scientists. Simulators are also developed exclusively for power systems. Internet technology based Power System Simulator (InterPSS) is a distributed and parallel simulation tool for optimizing and enhancing the design, analysis, and operation of power systems~\cite{khaitan2012high}. 

\subsection{Data Center Simulations}
In a cloud environment, virtual machines are run on top of fewer real physical hosts on the data centers. Resources are allocated to the virtual machines, based on the defined policies and algorithms. For energy and cost efficiency, virtual machines are migrated and consolidated across the hosts. Researches on optimizing the data center efficiency are done on simulation environments. Most of the cloud simulators offer the required functionality for data center simulations. However, representation of data centers and host allocation at the infrastructure-as-a-service level consume a lot of memory and processing power. Hence, specific data center simulators are developed to handle the limitations faced by cloud simulators in simulating the data centers.

\paragraph*{GDCSim:}
Green Data Center Simulator (GDCSim) is a data center simulator for energy-aware data center design and resource management techniques~\cite{gupta2011gdcsim}. It attempts to unify the simulation of management techniques with the representation of physical characteristics of the data center~\cite{gupta2011gdcsim}. While Computational Fluid Dynamics (CFD)~\cite{anderson1995computational} simulators characterize the thermal effects and airflow patterns which are optimized for the computation of energy consumption of the data centers, they do not offer a holistic design for a data center modeling. GDCSim provides iterative design and thermal analysis, which are not offered by Computational Fluid Dynamics simulators.

\paragraph*{GreenCloud:}
GreenCloud is a packet level simulator that simulates energy-efficient cloud data centers~\cite{greencloud}. The Network Simulator - ns-2, a real-time network emulation tool~\cite{mahrenholz2004real}, is extended for simulating energy-aware clouds. VM power management and migration in Internet Data Centers (IDC) are modeled, using Xen\footnote{\url{http://www.xenproject.org/}} as the VMM~\cite{liu2009greencloud}.

\paragraph*{DCSim:}
DCSim is an extensible data center simulator, that simulates data centers that host infrastructure-as-a-service cloud environments~\cite{dcsim}. MDCSim, GDCSim, and GreenCloud lack the ability to model virtualization. But DCSim provides virtualized data centers for multiple tenants. CloudSim handles the virtualization for data centers in a similar manner, whilst DCSim focuses more on transactional and continuous workloads.

\paragraph*{MDCSim:}
MDCSim is a multi-tier data center simulator~\cite{lim2009mdcsim}. Focus is given for each individual tier in a cluster of data centers, which was lacking in the other data center simulators. Power management and server configurations are two considerable use cases of MDCSim in simulating 3-tier or n-tier data center deployments.

\paragraph*{Comparison of Cloud Simulators:}
Cloud simulators have overlapping features, while some of the features are specific to only a few simulators. While some simulators are quite generic, others tend to be more focused. A comparison of three cloud simulators, CloudSim, SimGrid, and GreenCloud is presented by Table~\ref{table:comparisonC}. As presented by Table~\ref{table:comparisonC}, all three simulators considered can simulate cloud and grid systems. SimGrid and GreenCloud have the ability to execute packet-level simulations. Federated clouds can be modeled and simulated by CloudSim and SimGrid. Simulating peer-to-peer and MPI applications is possible with SimGrid, out of the box. GreenCloud focuses on modeling energy-aware computational resources and data centers. However, this can be achieved by CloudSim and SimGrid as well. While CloudSim is developed in Java, and SimGrid in C, GreenCloud is developed in both C++ and TCL (Tool Command Language). Hence, customizing the GreenCloud simulator framework or developing simulations on top of it, require knowledge of both languages. Application scheduling and higher level simulations are handled effectively by CloudSim and SimGrid, where GreenCloud focuses on lower level simulations such as data center simulations.

\begin{table}[!t]
\caption{Comparison of Cloud Simulators}
\label{table:comparisonC}
\begin{tabular}{|c||c| |c| |c|}
\hline

 & \textbf{CloudSim} & \textbf{SimGrid} & \textbf{GreenCloud}\\
\hline
\hline
\textbf{Programming} & & & C++ and TCL \\
\textbf{Language(s)} & Java & C & (Tool Command Language) \\
\hline
\hline
\textbf{User Interface} & Console & Console & Graphical \\
\hline
\hline
\textbf{Features} & & &\\
\hline
Grid Simulations & \checkmark & \checkmark & \checkmark \\
\hline
Cloud Simulations & \checkmark & \checkmark & \checkmark \\
\hline
Application Scheduling & \checkmark & \checkmark & X \\
\hline
Modeling Data Centers & \checkmark & \checkmark & \checkmark \\
\hline
Modeling Energy-Aware & & &  \\
Computational Resources & \checkmark & \checkmark & \checkmark \\
\hline
P2P Simulations & X & \checkmark & X \\
\hline
MPI Simulations & X & \checkmark & X \\
\hline
Packet-level Simulations & X & With ns-3 & With ns-2 \\
\hline 
Modeling and Simulation of & & &  \\
Federated Clouds & \checkmark & \checkmark & X \\
\hline 

\hline 
\end{tabular}
\end{table}

\section{MapReduce Frameworks and Simulators}
\label{sec:2ms}
MapReduce is a paradigm of parallel programming that lets applications to be distributed across multiple computer nodes and executed with map() and reduce() procedures, similar to the primitives of functional programming languages~\cite{dean2008mapreduce}. Due to their embarrassingly parallel nature, MapReduce programs can easily be distributed to execute on a cluster, grid, or a cloud. Originally developed in Google, MapReduce open source implementations became widespread across the research community and industry. Hadoop~\cite{white2009hadoop} offers the most popular open source MapReduce framework~\cite{dean2010mapreduce}. In-memory data grid platforms such as Infinispan and Hazelcast also offer MapReduce implementations.

\subsection{MapReduce Simulators}
As MapReduce applications and systems are developed with an increasing complexity, necessity to simulate MapReduce executions became apparent, in order to study their (and that of underlying algorithms') performance, efficiency, scalability, and resource requirements. Some of the MapReduce simulations were built from scratch, while some were developed on top of the existing simulation frameworks of cloud or network. MapReduce simulators are often built on top of the frameworks of the MapReduce implementation that they try to simulate, such as Hadoop. We will look at some of the MapReduce simulators below.

\paragraph*{MRPerf:}
MRPerf is a simulator of the MapReduce implementation of Hadoop, built using ns-2~\cite{wang2009using}. It identifies the potential performance bottlenecks that may be encountered in a MapReduce application running on top of a Hadoop platform. The simulator has been verified by its developers by comparing the results with the original MapReduce setups on Hadoop clusters. Job execution time, amount of the data transferred, and time taken for each phase of the job are output from the simulator~\cite{wang2009simulation}. 

\paragraph*{HSim:}
HSim, another Hadoop MapReduce simulator following the same design paradigm of MRPerf, claims to improve the accuracy of the MapReduce simulations for the complex Hadoop applications. While MRPerf is focused on network topologies which is validated using Terasort, search, and index, HSim focuses further on the behavior of Hadoop for the map and reduce operations, by having the implementation built on top of Hadoop~\cite{liu2013hsim}.

\paragraph*{SimMR:}
SimMR is a MapReduce simulator that can replay the tasks from the logs of the real workloads produced by Hadoop, executing the tasks within 5\% of the time the MapReduce task originally takes to execute~\cite{verma2011play}.

\paragraph*{Energy Aware MapReduce Simulations:}
Time sensitive interactive workloads for MapReduce tasks are defined as MapReduce with Interactive Analysis (MIA)~\cite{chen2012energy}. Since these MapReduce tasks are run on bigger clusters, energy becomes a very important concern to address. BEEMR (Berkeley Energy Efficient MapReduce) is a MapReduce workload manager that is energy efficient~\cite{chen2012energy}.

\subsection{Extending Cloud Simulators for MapReduce Simulations}
Extending cloud simulators for simulation of MapReduce avoids re-inventing the wheel, while exploiting the community and wisdom of the matured cloud and grid simulators. While some of the extensions are implemented by the researchers who originally developed the respective grid or cloud simulator, others are implemented by independent researchers. We will now look at the notable projects that provide MapReduce simulations, by extending known cloud simulators.

\paragraph*{CloudSimEx:}
CloudSimEx\footnote{\url{https://github.com/Cloudslab/CloudSimEx}} is a set of extensions to CloudSim, developed by a team of developers from the CLOUDS laboratory. Among the other features, CloudSimEx also possesses the ability to simulate MapReduce applications. CloudSimEx models MapReduce as a job composed of map tasks and reduce tasks. Map tasks and reduce tasks are depicted by the classes that extend the Task class, which is a sub class of Cloudlet. The class diagram of MapReduce in CloudSimEx is shown by Figure~\ref{fig:cloudsimex}.
\begin{figure}[ht]
\begin{center}
 \resizebox{0.3\columnwidth}{!}{
  \includegraphics[width=0.3\textwidth]{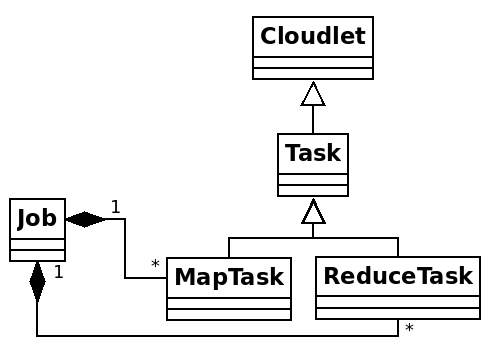}
 }
\end{center}
 \caption{CloudSimEx MapReduce Implementation}
 \label{fig:cloudsimex}
\end{figure}

CloudSimEx can also run multiple experiments in parallel, in different JVM processes. Users can spawn multiple independent JVM process, and redirect their standard outputs to a single place, with CloudSimEx. Due to the heavy use of static data in CloudSim, different threads cannot be used for independent experiments.

\paragraph*{MRSG:}
MRSG is a MapReduce Simulator built on top of SimGrid, providing APIs to prototype MapReduce policies and evaluate the algorithms~\cite{kolberg2013mrsg}. It was tested against the executions on Hadoop MapReduce framework to show the similarity between the real executions and simulations~\cite{kolberg2013mrsg}. As MapReduce tasks require more nodes to be simulated, MapReduce simulators should be built on top of the frameworks that are highly scalable, and capable of simulating more nodes. Built on top of SimGrid, MRSG has a higher scalability than the MapReduce simulators built on top of the other cloud and network simulator frameworks~\cite{kolberg2013mrsg}.

\subsection{MapReduce Over Peer-to-Peer Simulators}
MapReduce frameworks can be built on top of peer-to-peer frameworks, as peer-to-peer frameworks offer constructs for parallel processing and a huge number of nodes for the execution of map() and reduce() constructs. A parallel processing framework following the MapReduce paradigm has been built using Brunet peer-to-peer framework~\cite{lee2011parallel}.

Peer-to-peer network simulators such as PeerSim and PlanetSim have higher scalability as peer-to-peer networks require more nodes to be simulated~\cite{naicken2011towards}. Considering the number of nodes to be simulated as the means of scalability, PeerSim has the highest scalability, as it can simulate one million nodes~\cite{casanova2008simgrid}. Hence, a MapReduce simulator can exploit the inherent scalability provided by the peer-to-peer networks, though it requires further development to adopt the peer-to-peer simulators into MapReduce execution simulations. ChordMR is a recent MapReduce simulator that uses PeerSim as the core simulation engine~\cite{wu2014chordmr}. ChordMR core simulation engine does not have a coordinator or a central manager node, unlike the other MapReduce frameworks following the Hadoop and Google style design. Integration of MapReduce paradigm into a peer-to-peer framework gives more use cases such as counting the number of nodes in the overlay network, distributed data mining, and resource discovery of P2P grid~\cite{wu2014chordmr}. This highlights the synergistic effect gained from integrating MapReduce into the other frameworks and simulators initially intended for cloud or peer-to-peer networks.

We discussed cloud and MapReduce simulators in detail, while also discussing some specific simulators for data centers, peer-to-peer and overlay networks, and energy-aware deployments. The simulator platforms are converging, as simulators are extended beyond their original intended use case. However, the simulators often limit themselves to running on a single server or computer node as their execution platform. In the upcoming section, we will discuss the distributed execution frameworks that provide a unified view of a computer cluster, as a single larger server. Further discussions are on cycle sharing model, which enables sharing resources across different server instances, providing a desktop grid for large executions.

\section{In-Memory Data Grids}
\label{sec:2de}
Multiple Java in-memory data grids exist, both open source and commercial. In particular, two open source data grids, Hazelcast and Infinispan as well as two commercial platforms, Oracle Coherence and Terracotta BigMemory, are discussed below, as they are very representative cases by virtue of being widely used in application distribution.

In-memory data grids such as Hazelcast, Infinispan, and Coherence provide cache mechanism with cache APIs such as MemCached, JCache, and Ehcache. Implementations of distributed maps are provisioned as the storage elements. Search capabilities are provided for the objects stored in the grid through an SQL and predicate based APIs. Backups and recovery mechanism ensure a fault-tolerant data grid in memory.

The in-memory data grids can be configured with a persistent data store, providing a logical data view between the persistence layer and the application. When the queried data is available in the in-memory data grid, it is returned, and if not, it will be retrieved from the persistent store. Similarly, upon updates, data grid gets updated and propagates the updates to the data store immediately or later, based on the configured update policies. In-memory data grids themselves operate as key-value stores in-memory. Hence, having a data store is optional, and is required only when the application layer contains objects that would not fit into the shared in-memory grid provided by the cluster. In such cases, the objects are evicted from the data grid periodically, and loaded from the persistence store as required. Having multiple nodes configured with fail-over provides fault tolerance to the data grid. As an in-memory NoSQL store to replace the traditional databases, or a caching layer above persistent storage, data grids enhance the performance of the deployment architecture.

\subsection{Hazelcast}
Hazelcast provides distributed implementations for the java.util.concurrent package~\cite{hazelcast}. By extending concurrent hashmaps, executor service, and other data structures to function in a distributed environment, Hazelcast provides a seamless development and deployment of a distributed execution environment. Computer nodes running Hazelcast can join or create a Hazelcast cluster using either multicast or TCP-IP based join mechanisms. Additionally, Amazon web service instances with a Hazelcast instance running, can use the Hazelcast/AWS join mechanism to form a Hazelcast cluster with the other EC2 instances. Multiple Hazelcast instances can also be created from a single node by using different ports, hence providing a distributed execution inside a single machine. Hazelcast Management Center is a tool that monitors execution and distribution of the data structures across the partitions, as well as the status and health of the nodes. 

\paragraph*{Distributed Storage and Cache Management:}
Hazelcast supports consistent hashing. Hence partitioning appears uniform with minimal reshuffling of objects when a new instance joins in, or when an existing instance leaves the cluster. Data partition table is achieved for each distributed object by calculating $hash(key) \% partitionCount$ for their keys respectively, where $partitionCount$ is 271 by default. Owner of the key can be found for each of the key. Related objects can be stored together by controlling the partitions by defining the keys of the respective objects in the format of $key@partitionKey$.

As Hazelcast distributes the objects to remote JVMs, the distributed objects must be serializable, or custom serializers must be developed and registered for each of the classes that are distributed. Hazelcast custom serialization requires the classes to be serialized to have public setters and getters for the properties that should be serialized. If some of the properties of the class should not be serialized, they should be marked as $transient$, to inform the serializer to ignore them in the serialization. Unless the default constructor is used to initialize the objects that are serialized, the parameters of the constructor should be indicated in the serializer class. 

Hazelcast stores the distributed objects in $BINARY$, $OBJECT$, or $CACHED$ formats, as defined by the tag $<in-memory-format>$ in $hazelcast.xml$. $BINARY$ in-memory format is the default in-memory format that stores the objects in their serialized binary format. $OBJECT$ format, on the other hand, stores the simple objects in the deserialized form, hence eliminating the serialization costs. When having to store custom complex objects and access them remotely, the object should be stored in $BINARY$ format, where it can be stored in $OBJECT$ format if the distributed objects are always accessed from the same instance locally. Exploiting the partition-awareness of Hazelcast, related objects can be stored in the same instance, reducing the data transmission and remote invocations.

Hazelcast has caching properties which can be enabled using a $<near-cache>$ tag for the maps. When near-caching is enabled, it caches the objects to the instances that access the distributed objects frequently. $CACHED$ format exploits the cached value, hence increasing the performance using the local cached value of the remote objects. However, this increases the memory consumption by caching the objects accessed by the JVM locally, also reducing the consistency guarantees based on the update or invalidate policies of the cache entries. 

\paragraph*{Data Redundancy and Memory Management Policy:}
While synchronous backup makes a redundant copy of the data structure immediately as it is created or modified, asynchronous backups do it at a later time. Hence, asynchronous backups may be outdated. Synchronous backups enable an active replication, where the first response from any of the instances can be considered. However, asynchronous backups lead to a passive replication, where the primary instance holding the value should respond to the query. 

Near-cache makes Hazelcast to use the backup value of a distributed object that is stored in the instance itself. If outdated values are tolerated, asynchronous backups may be used to increase the performance of Hazelcast, with near-caching enabled, such that the asynchronously copied values in the same instance will be used in execution, reducing the remote invocations. Cache invalidation or update policies are used in favor of accuracy, to remove the cached up backups that are changed later in the original data object. Similarly, eviction policies remove the copies of distributed objects based on the defined eviction policies.

Hazelcast evicts the distributed object entries based on two eviction policies, Least Recently Used (LRU) and Least Frequently Used (LFU), as defined in the configuration properties. If an eviction policy is not defined, Hazelcast waits for the time out period to evict them. The time out period is based on the life time of the entries (time-to-live-seconds) and the time the entry stayed idle in the map (max-idle-seconds). These are by default infinite such that no entries are evicted though they are not used. This makes Hazelcast an in-memory key-value store that can function as an in-memory NoSQL storage. Persistence to disk can be achieved by configuring Hazelcast to use relational databases or NoSQL databases such as MongoDB\cite{chodorow2013mongodb}. This can be useful where the entire distributed objects cannot be stored in-memory due to their huge size, where eviction is configured along with the persistence layer of relational databases or NoSQL data stores.

\paragraph*{Execution Monitoring and Integration:}
Hazelcast Management Center is a tool from Hazelcast that is used to monitor the Hazelcast cluster and system health. This is deployed on a web container such as Apache Tomcat. It can monitor the local as well as the remote Hazelcast instances. It monitors the progress of distributed executors and the storage and state of the distributed objects. Management Center runs independently from the web container, separate from the Hazelcast clusters that it monitors. $hazelcast.xml$ the configuration file that is used to configure Hazelcast, is used to point to the running Management Center instance. Figure~\ref{fig:mancen} is a screen shot of Hazelcast Management Center 3.2 deployed into Apache Tomcat 8.0.5, monitoring an active Hazelcast cluster. Cluster groups can be secured with user name and password credentials in a shared network to prevent joins from unauthorized instances.
\begin{figure}[!htbp]
\begin{center}
 \resizebox{0.95\columnwidth}{!}{
  \includegraphics[width=0.95\textwidth]{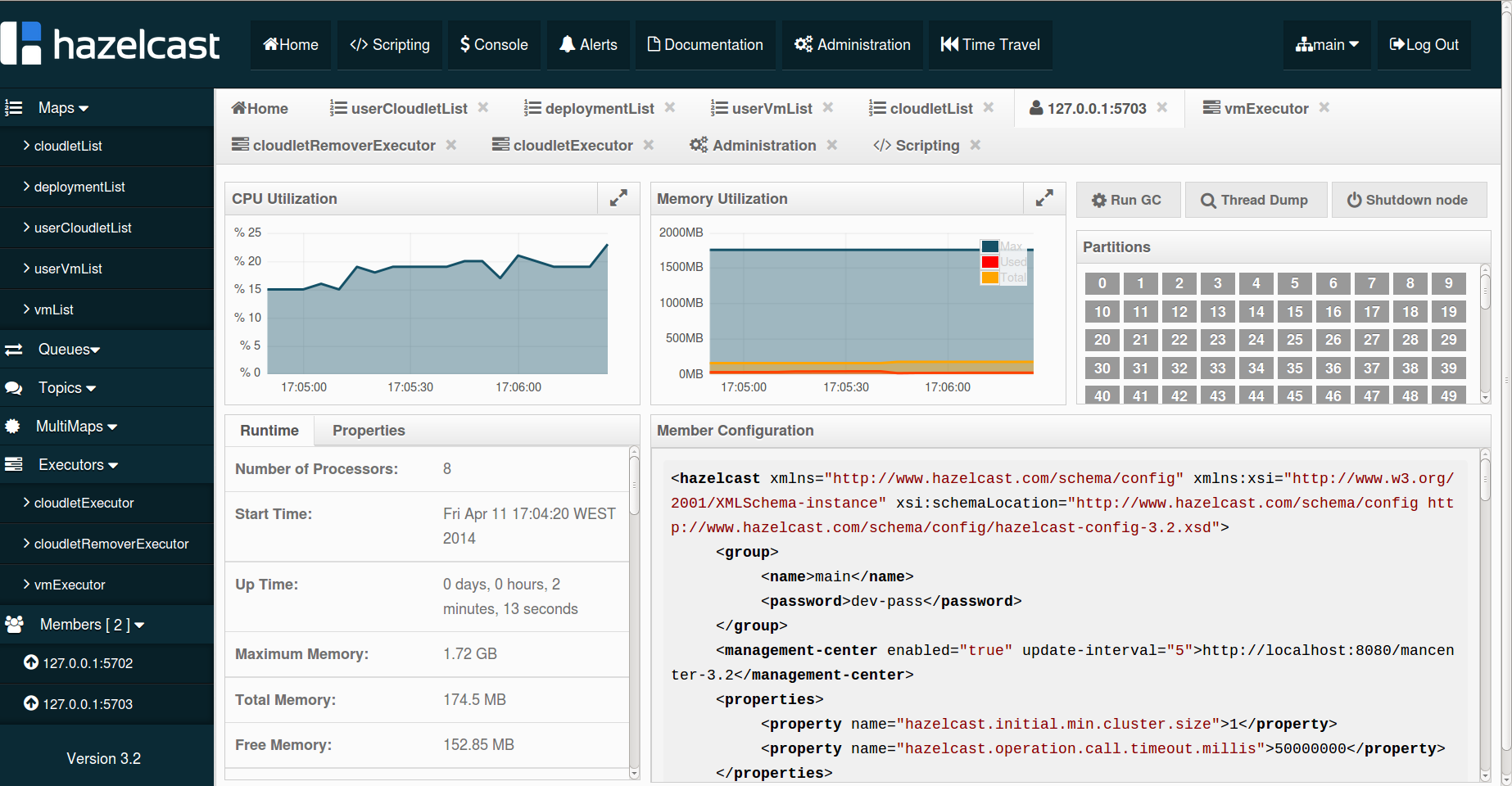}
 }
\end{center}
 \caption{Hazelcast Management Center Deployed on top of Apache Tomcat}
 \label{fig:mancen}
\end{figure}

\paragraph*{Hazelcast in Research:}
 There are distributed caching projects that exploit Hazelcast. Hazelcast has been already used in research, mostly to distribute the storage across multiple instances. A security research proposes performance improvements to Advanced Encryption Standards, using Hazelcast to distribute the storage of the objects~\cite{hazelsec}.

\subsection{Infinispan}
Infinispan is a distributed key/value data-grid~\cite{infinispan}. When used as a cluster-aware data-grid over multiple nodes, Infinispan can execute applications that would not run on a single node/computer due to the limited availability of resources. By utilizing multiversion concurrency control, Infinispan permits concurrent readers and writers, opposed to the coarse grained Java concurrency control and synchronization. Hence, when used as a local in-memory cache, Infinispan outperforms ConcurrentHashMap. While Infinispan can be used as a distributed cache for scaling the storage and execution out, fault-tolerance can be achieved with Infinispan as a replicated cache. 

Infinispan uses its own serialization, as default Java serialization is deemed as slow. For known types such as the internal objects such as commands and clustering messages of Infinispan cache, it uses a single byte magic number, instead of writing the complete class definitions to the stream. The magic numbers are registered with Infinispan using externalizers, provided by JBoss Marshalling\footnote{\url{http://www.jboss.org/jbossmarshalling}}. For distributed objects with custom data types that are developed by the users, Infinispan resort to the default Java serialization. Users can register externalizers for custom data types to make use of the efficient serialization.

\paragraph*{Infinispan in Researches:}
Infinispan has been used in many researches, as an in-memory data-grid. A self-adaptive middleware platform has been developed to provide transactional data access services, based on the in-memory data management layer of Infinispan~\cite{palmieri2012integrated}. Goal-oriented self-adaptive scalable platforms are researched using Infinispan as an in-memory persistence and cache solution~\cite{rosa2011goal}. Infinispan depends on two-phase commit based replication, which can further be made more efficient with partial replication techniques with weak consistency~\cite{ruivo2011exploiting}.

JBoss RHQ \footnote{\url{http://rhq.jboss.org/}} provides an enterprise management solution for Infinispan as well as the other projects from JBoss, which can be used to monitor the state and health of the Infinispan distributed cache instances. Infinispan offers JCache\footnote{\url{https://jcp.org/en/jsr/detail?id=107}} and MemCached\footnote{\url{http://memcached.org/}} cache APIs. Infinispan and Hazelcast have similar functionality, and both can be used as an in-memory cache. While Infinispan has been optimized to funtion as a distributed as well as a local cache, Hazelcast targets mostly to be a distributed cache. Hazelcast binary is a single jar with no external dependencies, where Infinispan needs a couple of jars to execute.

\subsection{Terracotta BigMemory}
Terracotta BigMemory is a commercial in-memory data management platform for big data. It is functionally comparable to Hazelcast, which provides an open source in-memory data grid. Terracotta Big-Memory 4.1 does not provide distributed executor service and MapReduce implementation, which are provided by Hazelcast 3.2 and Infinispan. A distributed execution framework for Terracotta has been developed by research~\cite{simao2013resource}.

Terracotta also develops and maintain the open source cache implementation, EhCache~\cite{wind2013instant}. Ehcache search also lets the users query and search the in-memory data. While replicated maps are not supported by Terracotta BigMemory, it is supported by Ehcache. Ehcache is complimentary to Terracotta BigMemory, in many of its features. As a platform for distributed memory, Terracotta has been used by researchers. $C^{2}Mon$ (CERN Control and Monitoring Platform) uses Terracotta BigMemory and Ehcache as the clustering and caching solutions to monitor complex scenarios~\cite{brager2014high}.

\subsection{Oracle Coherence}
As a commercial in-memory data grid, Oracle Coherence is used to distribute the backend systems such as databases and mainframes by fronting them~\cite{seovic2010oracle}. As Coherence allows loading data from multiple data sources to in-memory grid, it can be used as an integration solution. An existing application can be scaled using Coherence as an L2 cache. Coherence is released as standard, enterprise, and grid editions, where grid edition allows WAN networking and more sophisticated real time clients.

\paragraph*{Comparison of in-memory data grids:}
A comparison of Terracotta BigMemory 4.1 with the in-memory data grid platforms, Hazelcast 3.2, Infinispan 6.0, and Oracle Coherence 12.1.2 is presented by Table~\ref{table:comparison}. Terracotta is a server-client architecture, where the other data grids have a grid architectural topology. Distributed atomicity and concurrency are ensured through the distributed data structures such as distributed lock, distributed atomic long, distributed atomic reference, and distributed atomic semaphores, by Hazelcast, Terracotta, and Coherence. However, Infinispan does not offer these distributed atomic data structures.

Hazelcast also offers a multimap structure, where each key can contain multiple values, which is a feature that is not supported by the other data grids. Replicated maps with active and passive replication are supported by the data grids except Terracotta.

\begin{table}[!t]
\caption{Comparison of in-memory data platforms}
\label{table:comparison}
\begin{tabular}{|c||c| |c| |c| |c|}
\hline
&  &  & \textbf{Terracotta} & \textbf{Oracle}\\
 & \textbf{Hazelcast} & \textbf{Infinispan} & \textbf{BigMemory} & \textbf{Coherence}\\
\hline
\hline
\textbf{License} &  Apache &  Apache & Proprietary & Proprietary \\
& License v.2.0. & License v.2.0.&License & License\\
\hline
\hline
\textbf{Architectural topology} & Grid & Grid & Client-Server &Grid\\
\hline
\hline
Search & SQL API \& & SQL API \& & SQL API \& & SQL API \& \\
& Predicate API & Predicate API & Predicate API & Predicate API\\
\hline
\hline 
\textbf{Distributed Atomicity} & & &&\\
\hline
\hline
Lock \& Atomic Long & \checkmark& X & \checkmark & \checkmark\\
\hline
Atomic Ref \& Semaphore & \checkmark & X & \checkmark & X\\
\hline
\hline
\textbf{Cache} & \checkmark & \checkmark & \checkmark & \checkmark \\
\hline
Near Cache & \checkmark & \checkmark & \checkmark & \checkmark \\
\hline
Cache API & MemCached & MemCached & Ehcache & Coherence\\
& & and JCache &  &\\
\hline
\hline 
\textbf{Maps}& & &&\\ 
\hline 
Distributed Maps & \checkmark & \checkmark & \checkmark & \checkmark\\
\hline 
Replicated Maps & \checkmark & \checkmark & X & \checkmark\\
\hline 
Multimaps & \checkmark & X & X & X\\
\hline 
\hline 
\textbf{Distributed Execution} & & &&\\
\hline
Executor Service & \checkmark & \checkmark & X & X\\
\hline 
MapReduce & \checkmark & \checkmark & X & \checkmark\\
\hline 
\hline 
\textbf{Distributed Messaging} & & &&\\
\hline
Distributed Queue & \checkmark & X & \checkmark & \checkmark\\
\hline
Distributed Events & \checkmark & X & \checkmark & \checkmark\\
\hline 
\hline 
\textbf{Management} & & &&\\
\hline
Monitoring & \checkmark & \checkmark & \checkmark & X\\
\hline 
Management API & JMX and API & JMX & REST & JMX\\
\hline 
Backups & In-memory  & In-memory & Database& In-memory \\
 & replications & replications &  Backup Utility& replications \\
\hline
Recovery & \checkmark & \checkmark & \checkmark & \checkmark \\

\hline 
\hline 

Elasticity & \checkmark & \checkmark & X &\checkmark\\
\hline 
\end{tabular}
\end{table}

\paragraph*{Other In-Memory Data Grids:}
IBM WebSphere eXtreme Scale~\cite{arora2011scalable}, VMWare vFabric GemFire\footnote{\url{www.vmware.com/products/vfabric-gemfire/}}, Gigaspaces XAP~\cite{lwenstein2010benchmarking}, and Gridgain\footnote{\url{http://www.gridgain.com/}} are other notable in-memory data grids.

Linda is a programming model for parallel and distributed applications~\cite{wells2005coordination}. Concurrent model transformations are implemented on Gigaspaces XAP, using the Linda model~\cite{burgueno2013concurrent}. A cloud based document classification has been implemented using Gridgain~\cite{samovsky2012cloud,sarnovsky2013cloud}. An adaptive scheduling strategy has been developed for the cycle sharing models leveraging Gridgain, as a potential merge of in-memory data grids into desktop grids~\cite{reddy2012adaptive}. 
  %
  %
\section{Cycle Sharing}
\label{sec:2cs}
Cycle sharing or resource sharing is a model for acquiring resources from multiple computers during their idle time, for much complicated tasks that are heavy in computing and storage requirements. The resource sharing may be across a research lab among the computers in a shared network or a cluster, or across the globe. Volunteer computing is a public cycle sharing approach that acquires the computing resources from volunteers for a scientific, social, or an interesting common cause. The volunteer computing paradigm enables distributed execution of embarrassingly parallel problem on the private computers of geographically distributed volunteers. Specific CPU and memory intensive research areas have utilized the volunteer computing model, where millions of volunteers offer their computer resources, while they are idle. 

\subsection{Volunteer Computing}
\paragraph*{SETI@home:}
SETI (Search for Extraterrestrial Intelligence) is a project that seeks life outside the Earth. Presuming alien civilization who wish to communicate with the races in the Earth will use signals of very narrow band frequency to be easily distinguishable from the noise, signal processing is used to find out such signals standing out from the noise. This requires very high processing power as precision and accuracy of the analysis depends on the available processing power. Volunteer computing model of SETI@home replaces the supercomputers that processes the signals received for SETI to analyze for extraterrestrial intelligence~\cite{anderson2002seti}. 

\paragraph*{BOINC:}
BOINC (Berkeley Open Infrastructure for Network Computing) is an open source software that enables scientists to operate on public resource sharing model~\cite{anderson2004boinc}. It is a server, client, and statistics system, that is later used by SETI@home and other projects~\cite{beberg2009folding}. The projects work in different platforms, and respective clients are downloaded to the volunteers' computers to consume the resources. BOINC advices the volunteers to pick the projects based on their security practices and policies when consuming the volunteer resources, who owns the results from the computations, and whether the results are publicly available. Moreover, volunteers contribute to the projects that are interesting or useful to them.

Volunteers can pick the projects from a list\footnote{\url{http://boinc.berkeley.edu/projects.php}} of trusted projects endorsed and verified by BOINC, and volunteer their computing resources during their idle time to the projects. More BOINC-based projects are added to the list of projects. A more complete list of projects including those that are not endorsed by BOINC is also available\footnote{\url{http://boinc.berkeley.edu/wiki/Project_list}}. Some of the projects utilize the Graphics Processing Unit (GPU) on the volunteers' computers to execute their tasks much faster and more effectively.
 
Enabling Grids for E-SciencE (EGEE) is a service grid project aimed to build a large infrastructure for grid computing~\cite{laure2009enabling}. Service grids facilitate a two-way resource sharing among the participating entities, unlike desktop grids, where the clients installed in user computers merely contribute their idle cycles to the large projects hosted and managed centrally. However, the usage of service grids is limited to large projects due to the complexity of their operation and maintenance. EDGeS tries to merge the service grids with desktop grids, by enabling desktop grids such as BOINC to submit tasks to service grids~\cite{kacsuk2008towards}. BOINC extensions research further to utilize the volunteer computing model for other use cases, overcoming the limitations faced by BOINC. nuBoinc extends BOINC for a community cycle sharing, where users can provide tasks to be executed on other idling computers~\cite{silva2008nuboinc}. This model enables the users to be resource providers as well as task providers, opposed to the BOINC model of having the users only volunteering their resources. 

\paragraph*{Volunteer Computing in Research:}
Folding@home and Gnome@home attempts to find solutions to tackle the problems in computational biology using distributed computing model, with resources donated virtually over an idle time, by the volunteers~\cite{larson2002folding}. Executions of large biological simulations are partitioned and distributed across the resources. Communication overhead often exceeds the advantages of distributing the tasks that require frequent communication across the nodes. Folding@home algorithm effectively tackles this problem of parallelization and communication~\cite{shirts2006screen}. Folding@home is built using a client-server architecture, where its client connects with the server to get more work, and also informs the volunteer about the status of the jobs that are carried on~\cite{beberg2009folding}.

\subsection{Condor}
Condor is a scheduling system that maximizes the utilization of the workstations. Underutilized or idling workstations offer their computing resources to the workstations that are overloaded. This resource sharing increases the overall productivity of the research labs or the cluster of workstations~\cite{litzkow1988condor}. Failed tasks are scheduled to other workstations, such that all the tasks eventually complete. Condor has been integrated with other systems to provide solutions for more research problems. A centralized scheduler architecture is more vulnerable to failures, and is not scalable, though it can effectively decide which of the submitted jobs to be scheduled next, as all the idling workstations and tasks in the waiting queue are accessible to the centralized scheduler. Distributed scheduler architecture is resilient to failures as there is no central coordinating entity. However, it requires each of the requesting workstation to find an idling workstation on itself. Therefore, this is ineffective in scheduling, to find which of the tasks is assigned next. Hence, Condor uses a hybrid approach in resource management.

\paragraph*{Condor-B:}
Condor-B extends and integrates Condor and BOINC, such that volunteers can provide resources to a Condor pool using BOINC~\cite{kijsipongse2013scaling}.

\paragraph*{Condor-G:}
Condor-G~\cite{frey2002condor} extends Condor to use the intra-domain resource management protocols of the Globus Toolkit~\cite{foster2005globus}. Condor-G presents multi-domain resources as if they belong to a single domain~\cite{frey2002condor}. Due to the fact that the resources are from multiple domains, security becomes an important aspect to consider in Condor-G. Security and resource discovery and access are provided by Globus Toolkit~\cite{foster2005globus}.

\subsection{P2P Overlays for resource sharing}
While BOINC, Condor, and related extensions have central managing entities, peer-to-peer design is also exploited for a resource sharing for high throughput computing (HTC) without a central manager. Pools on Demand (PonD) is a middleware platform that provides a peer-to-peer overlay network for scalable resource discovery, creating a resource pool for large resource requirements~\cite{lee2012pond}. CCOF (Cluster Computing on the Fly) is a project for embarrassingly parallel applications with a master-slave model to utilize the idle cycles in a peer-to-peer network with fairness, such that the volunteers in turn receive idle cycles for the execution of their applications~\cite{zhou2004cluster}. GiGi presents gridlets as a computational workload that is semantic-aware, and implements a grid over an overlay network~\cite{veiga2007gigi}. Ginger is a middleware architecture to parallelize existing CPU-intensive applications without modifying the source code, exploiting the overlay networks~\cite{morais2011transparent}. 

The shift towards a peer-to-peer paradigm for the underlying architecture of the cycle sharing applications opens up more research avenues. Researches blur the demarcation of service and desktop grids and are expanding the reach of the cycle sharing and self-adaptive systems.

\paragraph*{Analysis:}
Simulations empower the researchers with an effective way of prototyping their research implementations. Multiple cloud and MapReduce simulators exist with different features. MapReduce simulators are independently built from scratch, or on top of cloud or network simulators. A comparison of features of the simulators show a performance/accuracy trade-off, which should be taken care of. Cloud and MapReduce simulators are unified as multi-purpose simulators from their intended use. 

However, the simulation frameworks still fail to exploit the existing distributed execution and desktop grid frameworks. No current cloud or MapReduce simulator is able to offer scale-out. Typically, the simulators are sequential, and run on a single computer, where computer clusters and in-memory data grids can be leveraged to execute larger simulations that cannot be executed on a single computer. Cycle sharing model can be utilized to provide means of sharing the resources across the simulation instances, allowing multiple independent simulations to execute in parallel, in a multi-tenanted way. Therefore, a scalable middleware platform for concurrent and distributed cloud and MapReduce simulations can be built, extending an existing cloud simulator, whilst exploiting the in-memory data grid platforms for an elastic environment, deploying an adaptive scaling strategy inspired by the volunteer computing model.

In-memory data grids and cycle sharing model provide resources for a distributed execution. They can be leveraged to execute larger simulations, to increase the performance of existing simulators, without sacrificing the performance. A potential merge of the simulation domain and distributed execution paradigm should be researched and implemented. Thus, even the simulations that can run on a single node can take advantage of more resources from the cluster, that it can run faster and more effectively. 

\paragraph*{Summary:}
In the first part of this Chapter, we discussed the cloud and MapReduce simulators. The latter part of the chapter discussed the distributed execution frameworks and desktop grids following the cycle sharing paradigm.

\chapter{Solution Architecture}
\label{chap:sa}
We will discuss the proposed solution in detail in this chapter. Section~\ref{sec:3con} presents the system and deployment architecture of the simulator, discussing in detail how the distribution of execution is achieved, and the partitioning strategies on how the simulation is partitioned and distributed across the instances. MapReduce simulator design, and how it functions in a multi-tenanted environment using Hazelcast or Infinispan, are also addressed. 

Dynamic scaling ensures an effective usage of the available resources. Instead of having all the available instances involved into a simulation task from the beginning, more instances can be involved adaptively, based on the load. Similarly, auto scaling ensures a cost-effective solution by scaling out based on the load, using Infrastructure-as-a-Service solutions such as Amazon EC2, or on private clouds such as Eucalyptus~\cite{nurmi2009eucalyptus}. We will further discuss dynamic scaling in Section~\ref{sec:3se}. Section~\ref{sec:3perf} reasons about and analyses the main aspects that will drive the speedup by the distributed execution of the simulations. Finally, Section~\ref{sec:3arch} will lead us into the detailed software architecture of the simulator.
  %
  %

\section{Concurrent and Distributed Middleware Platform}
\label{sec:3con}
As designed to run top of a cluster, $Cloud^{2}Sim$ attempts to execute larger and more complicated simulations that would not run on a single node, or consume a huge amount of time. A cluster of shared resources can be built over a cluster of computers, using the in-memory data grid frameworks. Simulations are executed on the cluster, utilizing the resources such as storage, processing power, and memory, provided by the individual nodes, as indicated by Figure~\ref{fig:datagrid}. Hazelcast and Infinispan are used as the in-memory data grid libraries in $Cloud^{2}Sim$. Based on the technical white papers, Hazelcast was chosen as the base platform to distribute the CloudSim simulations.

\begin{figure}[!htbp]
\begin{center}
 \resizebox{0.8\columnwidth}{!}{
  \includegraphics[width=0.8\textwidth]{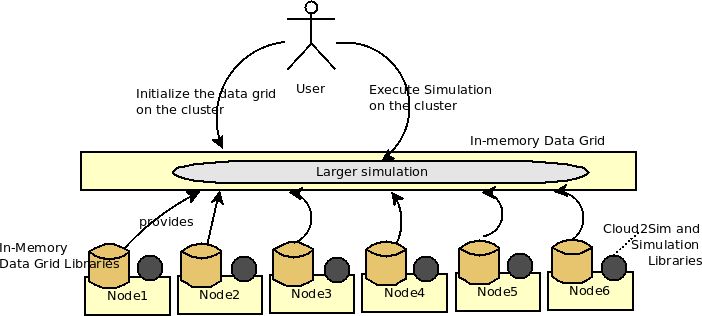}
 }
\end{center}

 \caption{High Level Use-Case of $Cloud^{2}Sim$}
 \label{fig:datagrid}
\end{figure}

$Cloud^{2}Sim$ functions in two basic modes as a concurrent and distributed simulator: cloud and MapReduce. It was decided to extend an existing cloud simulator to be concurrent and distributed, instead of writing a new cloud simulator from the scratch, to be able to take advantage of existing simulations. Developed as a Java open source project, CloudSim can be easily modified by extending the classes, with a few changes to the CloudSim core. Its source code is open and maintained. Hence, CloudSim was picked as the core simulation module to build the distributed simulator of this thesis. Cloud simulation further uses Hazelcast to distribute the storage of VM, Cloudlet, and Datacenter objects and also to distribute the execution, according to the scheduling, to the instances in the cluster. Users have the freedom to choose Hazelcast based or Infinispan based distributed execution for the cloud and MapReduce simulator, as the simulator is implemented on top of both platforms following the same design. Classes of CloudSim are extended and a few are also modified to be able to extend CloudSim with further functionality.  External dependencies such as Hazelcast and Infinispan are used unmodified, for added transparency and portability. The definition of cloud simulations and MapReduce simulations are independent by design. Cloud and MapReduce simulations can be executed independently, though experiments can be run utilizing both cloud and MapReduce simulations. 

\subsection{Partitioning of the Simulation}
\label{ssec:3partition}
As multiple instances execute a single simulation, measures are taken to ensure that the output is consistent as if simulating in a single instance, while having enhanced performance and scalability. Data is partitioned across the instances by leveraging and configuring the in-memory data grid. Each instance of the cluster executes part of the logic on the objects that are stored in the local partitions of the respective nodes. The logic that is distributed is simulations that are developed by the users. This includes the application-specific logic as well as the common system executions such as data center, VM, and cloudlet creation and scheduling.

Execution of simulations is improved, by leveraging the multi-core environments, and exploiting the multi-threaded programming. While CloudSim provides some means for a concurrent execution, its support is very limited. Simulations should be executed utilizing the multi-threaded environments, where the simulator itself runs the tasks concurrently, whenever that is possible and efficient. Runnables and callables are used to submit tasks to be run in a separate thread, while the main thread is executing its task. The relevant check points ensure that the threads have finished their execution and the values are returned from the callables, as required. 

A cluster can be formed by multiple instances. Multiple clusters can be used to execute parallel cloud or MapReduce simulations, as multiple tenants of the nodes. As each cluster is unaware of the other clusters, tenant-awareness is ensured so that the parallel experiments can be independent and secured from the other parallel simulations.

Pulling data from each of the nodes for execution has a higher communication cost. To overcome this, the data locality features provided for Hazelcast distributed executors are leveraged and used appropriately to send the logic to the data instead. Partition-awareness feature of Hazelcast is exploited in storing the distributed objects, such that the data that are associated with each other are stored in the same partition to decrease the remote invocations. 

Multiple partitioning strategies were implemented for different scenarios, as shown by Figure~\ref{fig:partitionapproaches}: \textit{Simulator - Initiator based strategy}, \textit{Simulator - SimulatorSub based strategy}, and \textit{Multiple Simulator instances strategy}. We will discuss each of these partitioning strategies further below.
\begin{figure}[!htbp]
\begin{center}
 \resizebox{0.8\columnwidth}{!}{
  \includegraphics[width=0.8\textwidth]{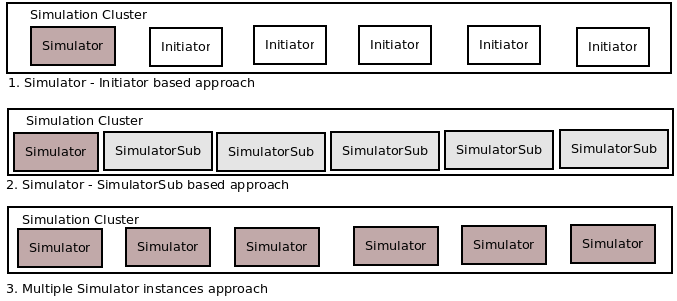}
 }
\end{center}
 \caption{Partitioning Strategies}
 \label{fig:partitionapproaches}
\end{figure}

\paragraph*{1. Simulator - Initiator based Strategy}
$Simulator$ is the complete $Cloud^{2}Sim$ with the simulation running. A Hazelcast instance is started by $Cloud^{2}Sim$ $Initiator$, which keeps the computer node connected to the Hazelcast cluster, offering the resources of the node to the data grid. The $Simulator$ instance is run from the master instance, where an instance of $Initiator$ is spawned from the other instances. Simulator acts as the master, distributing the logic to the Initiator instances. Part of the logic is executed in the master itself, and the execution is partitioned uniformly among all the instances, using the ExecutorService.

Pseudocode for the Initiator is presented in Algorithm~\ref{alg:initiator}.
\begin{algorithm}
  \caption{Initiator Instance}
\label{alg:initiator}
  \begin{algorithmic}
\State $initInstance()$
\While{ $(Simulation Is Executing)$ }
\State \textbf{Receive} Executor Service Executions \textbf{From} Master
\State \textbf{Execute} simulation step on the Data Partition
\EndWhile
\State $clearDistributedObjects()$ 
  \end{algorithmic}
\end{algorithm}

Pseudocode for the Master Instance is presented in Algorithm~\ref{alg:master}. The parts of the simulation that can be distributed include,
\begin{enumerate}[i]
  \item Creation of resources, cloudlets and VMs.
  \item Allocation of resources.
  \item Independent execution of the entities such as cloudlets, VMs, and brokers.
\end{enumerate}
Further simulation components that can be distributed depend on each simulation, and each simulation should ensure to partition the simulation space accordingly to maximize and optimize the parts of the execution that can be distributed. Distributed brokers and distributed counter-parts of cloudlets and VMs are designed, such that the application developer can utilize them to ensure smooth transitioning of the simulations to the distributed environment.

\begin{algorithm}
  \caption{Simulation Master Instance}
\label{alg:master}
  \begin{algorithmic}
\State Start and Initialize $Cloud^{2}Sim$ Cluster
\Repeat
\While{ $(SimulationIsDistributable)$ }
\State \textbf{Send} Executor Service Executions To Other Instances
\State \textbf{Execute} on the Data Partition
\State \textbf{Execute} Logic Partitions of its own
\EndWhile
\State \textbf{Process} Received Partitions from Other Instances
\State \textbf{Execute} Core Simulation That cannot be Distributed.
\Until $(SimulationIsCompleted)$
\State $presentSimulationOutput()$
\State $clearDistributedObjects()$ 
  \end{algorithmic}
\end{algorithm}

\paragraph*{2. Simulator - SimulatorSub based Strategy}
One instance contains the Simulator, which is the master, where others execute SimulatorSub, which are the slave instances. Master coordinates the simulation execution. Execution is started by all the instances and parts of the execution are sent by each instance respectively to the other instances, using the ExecutorService. Hence, the load on the master is reduced. Some of the unparallelizable tasks can be delegated to the primary worker, which is an instance other than the master instance, that is decided upon the cluster formation. This mitigates overloading the master instance.

The master instance still processes the core logic that cannot be distributed, and hence the pseudocode for the master instance does not change. Execution of SimulatorSub instances is described by Algorithm~\ref{alg:simulatorsub}.
\begin{algorithm}
  \caption{SimulatorSub Instances}
\label{alg:simulatorsub}
  \begin{algorithmic}
\Repeat
\While{ $(simulationIsDistributable)$ }
\State \textbf{Send} Executor Service Executions To Other Instances
\State \textbf{Execute} on the Data Partition
\State \textbf{Execute} Logic Partitions of its own
\EndWhile
\Until $(simulationIsCompleted)$
\State $clearDistributedObjects()$ 
  \end{algorithmic}
\end{algorithm}

\paragraph*{3. Multiple Simulator Instances Strategy}
There is no predefined Simulator master in this strategy. The instance that joins first becomes the master at run time, where other instances function as $SimulatorSub$ instances. Logic is partitioned across the instances using the partitioning algorithms defined in $Cloud^{2}Sim$ distributed data center brokers. $PartitionUtil$ manages the partitioning of the data and execution, manipulating the data structures across the instances. It provides the initial and final IDs of the data structure such as cloudlets and VMs, given the total number of the data structure elements and the initial offset. Figure~\ref{fig:partition} shows a higher level view of the partitioning. Here, the distributed storage is provided by all the physical nodes that host the instances in the execution cluster. Each type of distributed objects such as cloudlets and VMs are partitioned to be stored in the instances. Partition is tracked using the object IDs, where the initial and final IDs are marked. Logic is executed in the objects that are stored in the specific instance, minimizing the remote invocations.

\begin{figure}[!htbp]
\begin{center}
 \resizebox{0.7\columnwidth}{!}{
  \includegraphics[width=0.7\textwidth]{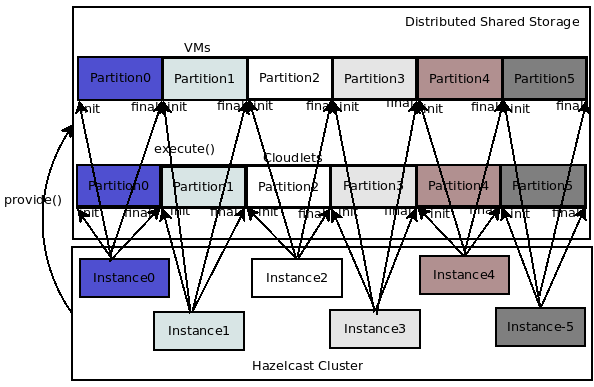}
 }
\end{center}
 \caption{Partition of storage and execution across the instances}
 \label{fig:partition}
\end{figure}

The \textit{Simulator - Initiator based Strategy} is chosen for implementation in tasks that are effectively scheduled by the single master to all the instances that are joined, such as the MapReduce simulator. The \textit{multiple Simulator instances strategy} is used in the CloudSim simulations such as the simulation of matchmaking-based application scheduling, where the simultaneous instances are more effective, than having a single static master that handles most of the task. The \textit{Simulator - SimulatorSub based strategy} is proposed for the compound simulations involving both Cloud and MapReduce executions, or simulating MPI workflows. The \textit{multiple Simulator instances strategy} is usually preferred over the \textit{Simulator - SimulatorSub based strategy} as it is easier to maintain since it does not fragment the logic, and also electing the master at run time is more effective in terms of scalability and fault-tolerance.

Existence of the master instance is always ensured in the \textit{multiple Simulator instances strategy}. The instance that joins the cluster as the first instance in the cluster becomes the master, where in the \textit{Simulator - SimulatorSub based strategy}, the instance of Simulator should be manually started before the sub instances, and this may become a bottleneck. Moreover, when backups are available, the \textit{multiple Simulator instances strategy} is resilient to failures as when the assigned master fails, another instance can take over as the master. This is not possible in the other strategies, as the master is chosen statically, and the other nodes do not contain the same code as the master instance.

\subsection{Multi-tenancy in $Cloud^{2}Sim$}
A multi-tenanted experiment executes over a deployment, composed of multiple clusters of (Hazelcast) instances, across multiple physical nodes. A tenant is a part of the experiment, represented by a cluster. An instance is attached to a single cluster, and is tenant-specific. Data specific to a tenant is stored in its instances of the cluster. The specific instance to store is defined by the $Cloud^{2}Sim$ design, leveraging and configuring the HazelcastInstanceAware and PartitionAware features to decide the optimal instance.

A coordinator node has instances in multiple clusters and hence enables sharing information across the tenants through the local objects of the JVM. Due to the 1:1 mapping between a cluster and a tenant, a tenant may not span across multiple clusters. This does not cause under-utilization, as multiple clusters can co-exist in and utilize the same nodes. Fault-tolerance is easily ensured by enabling synchronous backups, by just changing the configuration file. Thus, even if a node goes down, the tenants will not suffer.
\begin{figure}[ht]
\begin{center}
 \resizebox{0.6\columnwidth}{!}{
  \includegraphics[width=0.6\textwidth]{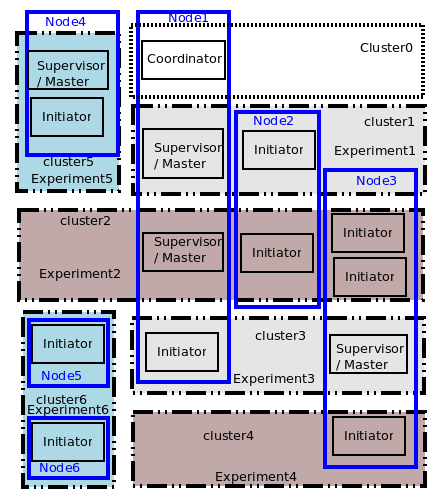}
 }
\end{center}
 \caption{A Multi-tenanted Deployment of $Cloud^{2}Sim$}
 \label{fig:mapreducecluster}
\end{figure}

Figure ~\ref{fig:mapreducecluster} depicts a sample deployment of 6 nodes configured into 7 clusters to run 6 experiments in parallel. Both cluster1 and cluster3 contain 2 nodes - the Master/Supervisor and one Initiator instance, running an experiment. Cluster2 contains 3 nodes, with the third node having 2 Initiator instances running. Cluster4 contains an instance of Initiator, ready to join a simulation job, when the Master instance joins. Cluster5 consists of node4, which hosts both Initiator and Master instances. Cluster6 contains node5 and node6, both running Initiator instances. Node1 hosts 2 Master instances, one in cluster1, and the other in cluster2. It also hosts an Initiator instance in cluster3. It has a Coordinator running on cluster0. Coordinator coordinates the simulation jobs running on cluster1 and cluster2 from a single point, and prints the final output resulting from both experiments or tenants. This is done externally from the parallel executions of the tenants, and enables a combined view of multi-tenanted executions.

Node - Experiment mapping can also be represented using a matrix notation of (Node X Experiment). The matrix for the multi-tenanted deployment depicted by Figure~\ref{fig:mapreducecluster} is given below.

Multi-tenanted Deployment =
\bordermatrix{\text{}&Exp_1&Exp_2& Exp_3 &Exp_4 & Exp_5 & Exp_6\cr
                Node_1&  S+C  & S+C & I & 0 & 0 & 0\cr
                Node_2& I & I & 0  &  0 & 0 & 0\cr
                Node_3& 0 & 2I & S & I & 0 & 0\cr
                Node_4& 0 & 0 & 0 & 0 & S + I & 0 \cr
                Node_5& 0  & 0 & 0 & 0 & 0  & I\cr
                Node_6& 0  & 0 & 0 & 0 & 0  & I}

Here, $S$ represents Supervisor/Master, $I$ represents Initiator, and $C$ represents coordinator. Though a deployment is better represented with nodes in horizontal and experiments/clusters in vertical, Figure~\ref{fig:mapreducecluster} is drawn with clusters in horizontal, for a better clarity of the representation.

\paragraph*{Cloud Simulations:}
$Cloud^{2}Sim$ is designed on top of CloudSim, where cloud2sim-1.0-SNAPSHOT can be built using Maven independently without rebuilding CloudSim. Modifications to CloudSim are very minimal. $Cloud^{2}Sim$ enables distributed execution of larger CloudSim simulations. The compatibility layer of $Cloud^{2}Sim$ enables the execution of the CloudSim simulations with minimal code change, on top of either the Hazelcast and Infinispan based implementations, or the pure CloudSim distribution, by abstracting away the dependencies on Hazelcast and Infinispan, and providing a compatible API. 

\paragraph*{MapReduce Simulations:}
Design of the MapReduce simulator is based on a real MapReduce implementation. A simple MapReduce application executes as the Simulator is started. The number of times map() and reduce() are invoked can easily be configured. The MapReduce simulator is designed on two different implementations, based on Hazelcast and Infinispan, making it possible to benchmark the two implementations against each other. Multiple simulations are executed in parallel, without influencing others, where an instance of a coordinating class could collect the outputs from the independent parallel MapReduce jobs carried out by different clusters. 

\section{Scalability and Elasticity}
\label{sec:3se}
$Cloud^{2}Sim$ achieves scalability through both static scaling and dynamic scaling. Static scaling is the scenario where $Cloud^{2}Sim$ uses the storage and resources that are initially made available, when instances are started and joined manually to the execution cluster. Multiple nodes can be started simultaneously at the start-up time for large simulations that require large amount of resources. $Initiator$ instances can also be started manually at a later time, to join the simulation that has already started. Simulations begin when the minimum number of instances specified have joined the simulation cluster. $Cloud^{2}Sim$ scales smoothly as more Hazelcast instances join the execution. 

Scaling can also be achieved by $Cloud^{2}Sim$ itself dynamically without manual intervention, based on the load and simulation requirements. When the load of the simulation environment goes high, $Cloud^{2}Sim$ scales itself to handle the increased load. Dynamic scaling of $Cloud^{2}Sim$ provides a cost-effective solution, instead of having multiple instances being allocated to the simulation even when the resources are under-utilized.

Since scaling introduces the possibility of nodes joining and leaving the cluster, as opposed to the static execution or manual joins and exits of instances, scalable simulation mandates availability of synchronous backup replicas, to avoid losing the distributed objects containing the simulation data upon the termination of an instance. 

A health monitor was designed to monitor the health of the instances, and trigger scaling accordingly. The health monitoring module runs from the master node and periodically checks the health of the instance by monitoring the system health parameters such as the process CPU utilization, system CPU utilization, and the load average. Based on the policies defined in the configuration file, the health monitor triggers the dynamic scaler. When the current observed value of the monitored health parameter (such as load average or process or system CPU utilization) is higher than the $maxThreshold$ and the number of total instances spawned is less than the $maxInstancesToBeSpawned$, a new instance will be added to the simulation cluster. Similarly, when the current observed value is lower than the $minThreshold$, an instance will be removed from the simulation cluster. Pseudocode for the dynamic scaling based on health monitoring is presented in Algorithm~\ref{alg:elasticity}.

\begin{algorithm}
  \caption{Dynamic Scaling}
\label{alg:elasticity}
  \begin{algorithmic}
\While{ $(TRUE)$ }
\State $getCurrentSystemHealthStatus()$
\If{$(load \geq maxThreshold$ \textbf\\{AND} $currentlySpawnedInstances < maxInstancesToBeSpawned)$}
    \State $scaleOut()$\Comment{add instance}
    \State $wait(timeBetweenScaling)$
\ElsIf{$(load\le minThreshold)$}
    \State $scaleIn()$\Comment{remove instance}
    \State $wait(timeBetweenScaling)$
\Else
    \State $wait(timeBetweenHealthChecks)$
\EndIf
    \EndWhile  
  \end{algorithmic}
\end{algorithm}

During scale out, more instances are included into the simulation cluster, where scale in removes instances from the simulation cluster, as the opposite of scale out. Dynamic scaling is done in two modes - auto scaling and adaptive scaling, as discussed below.

\subsection{Auto Scaling}
By default, the $Cloud^{2}Sim$ auto scaler spawns new instances inside the same node/computer. The auto-scaling feature is available out of the box for Hazelcast paid/enterprise versions. As $Cloud^{2}Sim$ uses the free and open source version of Hazelcast, auto scaling feature is designed on top of Hazelcast, using the health monitoring module of $Cloud^{2}Sim$. 

When there is only a limited availability of resources in the local computer clusters that is insufficient to simulate a large scenario, $Cloud^{2}Sim$ can be run on an actual cloud infrastructure. Hazelcast can be configured to form a cluster on Amazon EC2 instances, with the Hazelcast instances running on the same AWS\footnote{\url{https://aws.amazon.com/}} account. When using AWS join mechanism provided by Hazelcast to form the cluster, Hazelcast uses the access key and secret key to authorize itself into forming the cluster. If no AWS security group is mentioned, all the running EC2 instances will be tried, where mentioning a security group will limit the search to only the instances of the same security group. Ports that are involved in Hazelcast clustering should be open and permitted in the EC2 instances. Scaling can be triggered by the $Cloud^{2}Sim$ health monitoring or using the scaling policies configured with AWS Auto Scaling and Amazon Cloud Watch, as shown by Figure~\ref{fig:aws}.

\begin{figure}[!h]
\begin{center}
 \resizebox{0.6\columnwidth}{!}{
  \includegraphics[width=0.6\textwidth]{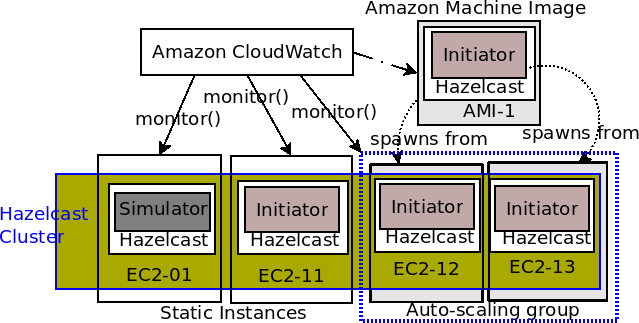}
 }
\end{center}
 \caption{Cloud Simulations on Amazon EC2 instances}
 \label{fig:aws}
\end{figure}

\subsection{Adaptive Scaling}
Adaptive Scaling is a scenario, where in a clustered environment, more computer nodes will be involved in an application execution based on the load. More instances will be attached to the simulation cluster when the load is high, and instances will be detached or removed from simulation when the load is low. We will discuss two of the design approaches that were considered, as they appear to be the logical and more straight-forward options, and will further elaborate why they were impossible without modifying Hazelcast. The final and feasible design without requiring code modifications to Hazelcast is presented after the two failed alternatives.

\paragraph*{1. Pause and Resume approach:}
Pause and resume instances, within a single cluster, which is not available out of the box in Hazelcast.

\paragraph*{2. Group Migration approach:}
In this approach, the deployment has two groups/clusters - cluster-main and cluster-sub. Instances from a cluster know nothing about those in the other clusters. Instances in cluster-sub are basically stand-by, where the cluster-main does all the simulation work. When the cluster-main group is overloaded, more instances from cluster-sub will be added to the group cluster-main and removed from cluster-sub. When the load goes down, they will be moved back to cluster-sub. 

Hazelcast does not indicate all the running instances. $Hazelcast.getAllHazelcastInstances()$ merely provides a list of instances running inside a single JVM. $hazelcastInstance.getCluster().getMembers()$ gives the members of any given cluster. Considering these limitations, to be able to communicate with both groups, two Hazelcast instances are started in the master node - One is of the group cluster-main and a $middleMan Instance$ of the group cluster-sub. But it is not possible to change the group configuration of an instance at run time, and change its group (move from one group to another) programmatically. Hence, this design approach of migrating an instance from a group to another became infeasible in Hazelcast.

\paragraph*{3. Scaling Decisions in a separate cluster - IntelligentAdaptiveScaler approach:}
In this approach, the health monitor in the main instance monitors the load and health status of the main instance with simulation running in $cluster-main$, and shares this information with the $AdaptiveScalerProbe$ thread in $cluster-sub$, using the local objects, as they are from the same JVM. $AdaptiveScalerProbe$ shares this information with $IntelligentAdaptiveScaler$ (IAS) instances, which are threads from all the other nodes that are connected to $cluster-sub$. 

When IAS from one instance notices the high load in the master, it spawns an Initiator instance in the $cluster-main$, and sets the flag to false to avoid further scaling outs/ins. Monitoring for scaling out happens when there is no Initiator instance in the node, and monitoring for scaling in happens when there is an Initiator instance, for each individual node. This ensures 0 or 1 of Initiator instances in each node, and avoids unnecessary hits to the Hazelcast distributed objects holding the health information. Since IAS is in a separate cluster (cluster-sub) from the simulation (cluster-main), the executions are independent. 

This design was chosen for the implementation of the adaptive scaler, as it is the most feasible implementation choice. Pseudocode for $AdaptiveScalerProbe$ is further presented in Algorithm~\ref{alg:asp}, and Algorithm~\ref{alg:ias} presents $IntelligentAdaptiveScaler$.
\begin{algorithm}[h]
  \caption{Adaptive Scaler Probe Algorithm}
\label{alg:asp}
  \begin{algorithmic}
\Procedure {addInstance}{}
\State $toScaleOut \gets TRUE$ \Comment{Atomic Boolean}
\EndProcedure
\Statex
\Procedure {removeInstance}{}
\State $toScaleIn \gets TRUE$ \Comment{Atomic Boolean}
\EndProcedure
\Statex
\Procedure {probe}{}
\While{ $(TRUE)$ }
\State $wait(timeBetweenHealthChecks)$
\If{$toScaleOut$}
\State $toScaleOut \gets FALSE$
\State $nodeHealth.toScaleOut \gets TRUE$ \Comment{Distributed Map Entries}
\State $nodeHealth.toScaleIn \gets FALSE$
\ElsIf{$toScaleIn$}
\State $toScaleIn \gets FALSE$
\State $nodeHealth.toScaleIn \gets TRUE$
\State $nodeHealth.toScaleOut \gets FALSE$
\EndIf
    \EndWhile  
\EndProcedure
  \end{algorithmic}
\end{algorithm}

\begin{algorithm}[h]
  \caption{Intelligent Adaptive Scaler Algorithm}
\label{alg:ias}
  \begin{algorithmic}
\Procedure {initHealthMap}{} \Comment{During the start up}
\State $key \gets 0$
\If{$(nodeHealth.toScaleOut = NULL)$}
\State $nodeHealth.toScaleOut \gets FALSE$
\EndIf
\If{$(nodeHealth.toScaleIn = NULL)$}
\State $nodeHealth.toScaleIn \gets FALSE$
\EndIf
\EndProcedure
\Statex
\Procedure {probe}{}
\While{ $(TRUE)$ }
\State $wait(timeBetweenHealthChecks)$
\If{$(instances.count() = 0)$}

\If{$(nodeHealth.toScaleOut)$}
\State $nodeHealth.toScaleOut \gets FALSE$ \Comment{Set to false, before the atomic decision}
\State $Atomic\{$ \Comment{Distributed atomic flag}
\State $currentValue \gets key$
\State $key \gets 1$
\State$\}$
\If{$(currentValue = 0)$} \Comment{No recent scaling by any instances}
\State $spawnInstance()$
\State $wait(timeBetweenScalingDecisions)$
\State $key \gets 0$ \Comment{Cluster may scale again now}
\EndIf
\EndIf

\ElsIf{$(nodeHealth.toScaleIn)$}
\State $nodeHealth.toScaleIn \gets FALSE$
\State $Atomic\{$
\State $currentValue \gets key$
\State $key \gets -1$
\State$\}$
\If{$(currentValue == 0)$}
\State $shutdownInstance()$
\State $wait(timeBetweenScalingDecisions)$
\State $key \gets 0$
\EndIf
\EndIf

    \EndWhile  
\EndProcedure
  \end{algorithmic}
\end{algorithm}
 
\subsection{Elastic Deployments}
Adaptive Scaling is used to create prototype deployments with elasticity. Adaptive scaling is built as shown by Figure~\ref{fig:IAS}. When the simulations complete, the Hazelcast instances running in the cluster-main will be terminated, and the distributed objects stored in the cluster-sub will be cleaned. These instances just require Hazelcast and the adaptive scaler thread to keep them connected, providing their CPU and storage for the  simulation work voluntarily, in a BOINC-like cycle sharing model. The entire simulation code can be loaded and kept only on the master and exported transparently to other nodes joining it, and execute from all the nodes, following the \textit{Simulator - Initiator based Strategy}. All the member nodes are from the same network, that they have joined by TCP-IP or multicast. Hence the cycle sharing model of $Cloud^{2}Sim$ is not public as in voluntary computing. Due to this nature, the security implications involved in voluntary computing are not applicable to $Cloud^{2}Sim$.

\begin{figure}[!htbp]
\begin{center}
 \resizebox{0.6\columnwidth}{!}{
  \includegraphics[width=0.6\textwidth]{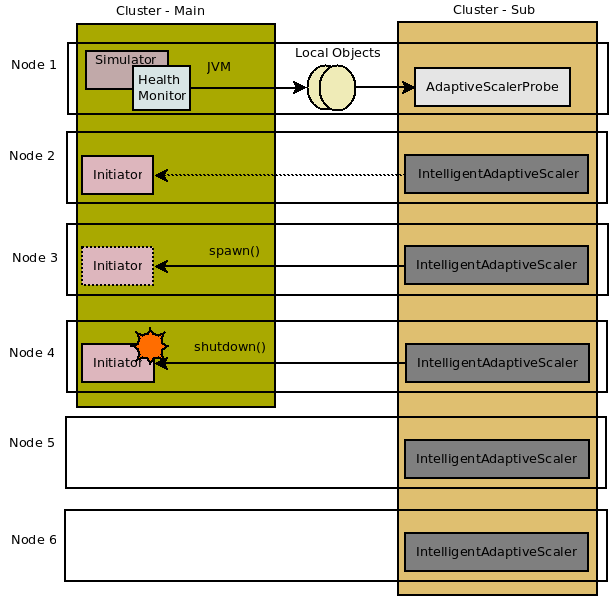}
 }
\end{center}
 \caption{Deployment of the Adaptive Scaling Platform}
 \label{fig:IAS}
\end{figure}

The scaling decision flag should be get and set in a concurrent and distributed environment atomically, ensuring that exactly one instance takes action of it. Access to the object that is used as the flag must be locked during update from any other instance in the distributed environment. 

\paragraph*{Parallel and Independent Simulations:}
Multiple Hazelcast clusters can be run from a single computer cluster or even a single machine. By exploiting this feature, multiple experiments can be run on $Cloud^{2}Sim$ in parallel, as different clusters are used for independent simulations. Different simulations are initialized from the same node, as shown below.

{\fontsize{10}{10}\selectfont
\begin{lstlisting}
String clusterGroup = HzConstants.MAIN_HZ_CLUSTER + id;

// Initialize the CloudSim library
HzCloudSim.init(ConfigReader.getNoOfUsers(), calendar, trace_flag, clusterGroup);
\end{lstlisting}
}
The adaptive scaling solution is further extended to have the node cluster providing its resources to different applications or simulations running on different Hazelcast clusters. Figure~\ref{fig:XIAS} shows the execution of two independent simulations in a cluster with adaptive scaling. The adaptive scaler functions as a $Coordinator$ instance, coordinating and allocating its resources to multiple tenants. Here, instead of representing the scaling decisions using single keys, distributed hash maps are used, mapping the scaling decisions and health information against the cluster or tenant ID. Similarly, the pointers to the master instances are mapped against the cluster ID, making it possible to refer to and coordinate multiple tenants from the coordinator.

\begin{figure}[!htbp]
\begin{center}
 \resizebox{0.8\columnwidth}{!}{
  \includegraphics[width=0.8\textwidth]{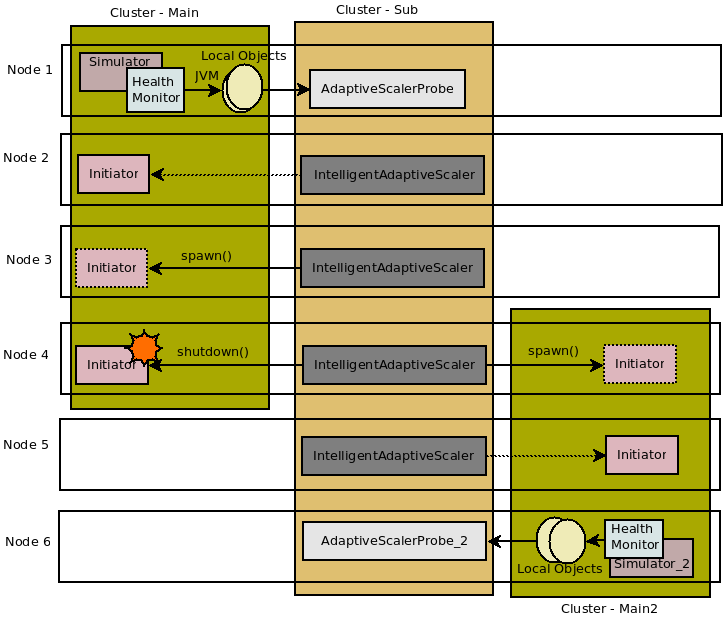}
 }
\end{center}
 \caption{An Elastic Deployment of $Cloud^{2}Sim$}
 \label{fig:XIAS}
\end{figure}

\section{Analysis of Design Choices Regarding Speedup and Performance}
\label{sec:3perf}
Speedup in a distributed or grid environment is a measure to compare how fast is the distributed application on multiple instances compared to its serial version on a single instance~\cite{hoekstra2005introducing}. Speedup in $Cloud^{2}Sim$ measures how fast the distributed simulation executes on multiple Hazelcast instances, compared to the time taken to execute the original simulation on CloudSim. Since Hazelcast initialization can be done just once for a series of simulations and experiments, we ignore the initialization time from the calculations.
\begin{equation} \label{eq:initequation}
T_{n} = \frac{k.T_{1}}{n} + (1-k).T_{1} + S + C + \gamma + F - \theta
\end{equation}

Here,\\
n - number of instances.\\
$T_{n}$ - Time taken by the distributed simulation on n instances.\\
k - Fraction of the code that should be run in a single instance and should not be distributed.\\
C - Latency or communication costs.\\
$\gamma$ - Coordination costs.\\
S - Serialization/deserialization costs.\\
F - Fixed costs.\\
$\theta$ - Performance gain from the inherently abundant resources of the data grid.

The fixed costs, F includes costs such as the time taken to initialize the threads, distributed executor framework, and other distributed data structures. It is required to serialize and deserialize the distributed custom objects such as cloudlets and VMs, as the objects are stored in remote locations. The distributed objects are serialized and stored in a binary form, and composed back to the respective objects upon deserialization. The memory and CPU cost from the serialization and deserialization is an immediate cost imposed by the integration of Hazelcast into $Cloud^{2}Sim$, which is present even when a single instance exists in the execution cluster. This overhead is common to all the distributed execution libraries, such that even the alternatives such as Infinispan imply this overhead. Serialization cost depends on the size or scale of the distributed simulation objects, where more objects to be serialized imposed more serialization cost. Hence, serialization cost is defined as,
\begin{equation} \label{eq:sc}
S = f_{1}(s)
\end{equation}
Here,\\
s - Size of the distributed simulation objects.\\

Coordination cost is defined as the cost caused by the instances coordinating themselves in the cluster. Communication costs and coordination costs increase with the number of instances, and with the latency caused by the physical distance between the nodes containing the instances, as well as the communication medium strength. As the number of instances is increased, these costs increase further, where they do not appear in the scenario of a single instance. If the same program runs in a single instance, communication cost and coordination cost become zero, where distributing it to more and more instances increases these. While communication cost depends on the size of the simulation, coordination cost is independent on the simulation, as it is caused by the framework regardless of the application that is running on top of it. Hence, communication cost can be defined as,
\begin{equation} \label{eq:cc}
C = f_{2}(n, d, w, s)
\end{equation}
Here,\\
d - Distance between the instances.\\
w - Bandwidth.\\
s - Size of the distributed simulation objects.\\

Coordination cost can be defined as,
\begin{equation} \label{eq:cd}
\gamma = f_{3}(n, d, w)
\end{equation}

As more nodes are introduced into the simulation cluster, application space is broken down into multiple nodes. Each node has to provide resources to only a fraction of the complete application. Hence, when more resources are made available for the simulation, larger simulations can be executed, while optimizing the cache of the computer. Moreover, this also increases the usage of memory, where large objects can be stored inside the memory, avoiding potential paging of larger objects, while minimizing memory contention. Hence, 
\begin{equation} \label{eq:theta}
\theta = f_{4}(N) 
\end{equation}
Here,\\
N - Number of physical nodes involved in the experiment.\\

Considering Equation~\ref{eq:sc}, Equation~\ref{eq:cc}, Equation~\ref{eq:cd}, and Equation~\ref{eq:theta}, Equation~\ref{eq:initequation} can be presented as,
\begin{equation} \label{eq:iequation}
T_{n} = \frac{k.T_{1}}{n} + (1-k).T_{1} + f_{1}(s) + f_{2}(n, d, w, s) + f_{3}(n, d, w) + F - f_{4}(N)
\end{equation}
 
$Speedup$, 
\begin{equation} \label{eq:baseequation}
S_{n} = \frac{T_{1}}{T_{n}}
\end{equation}
where e$fficiency$ is defined as,
\begin{equation} \label{eq:basefequation}
E_{n} = \frac{S_{n}}{n} = \frac{T_{1}}{n.T_{n}}
\end{equation}

While $speedup$ measures how faster the application performs with increasing number of instances, $efficiency$ provides a normalized measurement by dividing the speedup value with the number of instances. Hence, $efficiency$ may be used as a more realistic measure to find the number of instances to be involved in any execution for a cost-effective scaling.
From equations \ref{eq:iequation} and \ref{eq:baseequation}, speedup can be formulated as below,
\begin{equation} \label{eq:finalequation}
S_{n} = \frac{T_{1}}{\frac{k.T_{1}}{n} + (1-k).T_{1} + f_{1}(s) + f_{2}(n, d, w, s) + f_{3}(n, d, w) + F - f_{4}(N)}
\end{equation}

Percentage improvement (P) in performance can be presented as,
\begin{equation} \label{eq:percentage}
P = (1 - \frac{1}{S_{n}}) * 100%
\end{equation}

From Equation~\ref{eq:finalequation},
\begin{equation} \label{eq:percentageFinal}
P = (\frac{k.T_{1}(1-\frac{1}{n}) + \theta - S - C - \gamma - F}{T_{1}}) * 100\%
\end{equation}

Communication cost is minimized by the partition-awareness and data locality of the design. Objects are kept in deserialized object format, when they are in simple format and are accessed locally. Objects are serialized, and stored in binary format, when they are distributed. However, serialization cost is inevitable, and does not increase with the number of instances, unlike communication costs. Hence, we may expect increased execution time when 2 nodes are used, when the serialization cost is high, followed by a speedup when executing in more nodes.

If all the Hazelcast or Infinispan instances reside inside a single computer, latency will be lower. While this increases the speedup of simulations that can run on a single computer, applications with high resource requirements will suffer from lack of memory or CPU, as all the instances should share the resources of the computer. Usually, it is expected that the instances run on different computers on a research lab cluster. Though the communication cost will be lower than the geo-distributed cluster, as all the nodes in the cluster are co-located, it will be higher than running the instances inside a single computer. 

Availability of abundant resources speed up the executions that would not run effectively on limited resources provided by a single computer. Performance gain by distributed execution $\theta$ depends on the requirements of the simulations as well as the availability or lack of the resources in the nodes contributing to the cluster. Hence, for an embarrassingly parallel application with lower memory and processing requirements, distributing the load inside a single computer using multiple in-memory data grid instances may be more effective, where a more resource-demanding simulation will perform better on a cluster with multiple physical nodes.

\section{Software Architecture and Design}
\label{sec:3arch}
Distributed storage and execution for CloudSim simulations is achieved by exploiting Hazelcast. Infinispan integration with the compatibility layer ensures easy integration of Infinispan to replace Hazelcast as the in-memory data grid for CloudSim simulations. Figure~\ref{fig:cloud2simArch} depicts a layered architecture overview of $Cloud^{2}Sim$, hiding the fine architectural details of CloudSim. 
\begin{figure}[!h]
\begin{center}
 \resizebox{0.55\columnwidth}{!}{
  \includegraphics[width=0.55\textwidth]{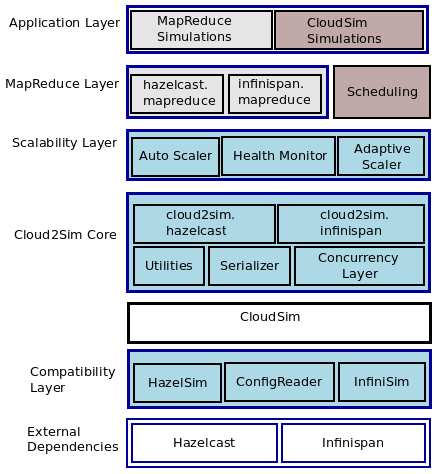}
 }
\end{center}

 \caption{$Cloud^{2}Sim$ Architecture}
 \label{fig:cloud2simArch}
\end{figure}

\subsection{$CloudSim$ Simulations}
As extending CloudSim, $Cloud^{2}Sim$ provides an API compatible with CloudSim, for the cloud simulations. Classes of CloudSim are extended as shown by Table~\ref{table:inheritance}, while preserving the invocation interfaces and code generality. This ensures easy adaptation and migration of $CloudSim$ simulations to $Cloud^{2}Sim$. Respective data structures of the CloudSim simulations can be easily ported to $Cloud^{2}Sim$ by using the extended classes as shown by Table~\ref{table:inheritance}, instead of the base CloudSim classes. By using bytecode enhancement, source code replacement/augmentation, and using object factory methods that can be intercepted or redefined, $CloudSim$ simulations can be executed on top of $Cloud^{2}Sim$, without the need to modify the simulation code.

\begin{table}[!t]
\caption{$Cloud^{2}Sim$ and CloudSim}
\label{table:inheritance}
\begin{tabular}{|c||c| |c|}
\hline
 & \textbf{Extended} & \\
\textbf{$Cloud^{2}Sim$ Class} &\textbf{CloudSim class} & \textbf{Core Responsibilities}\\
\hline
HzCloudSim & CloudSim & * Core class of the Simulator\\
& & * Initializes distributed data structures \\
\hline
HzDatacenterBroker & DatacenterBroker & * Implements distributed scheduling\\
\hline
 & - & * Starts Simulation based on the configuration \\
Cloud2SimEngine & & * Starts supportive threads \\
& & for scaling and health monitoring\\
\hline
PartitionUtil & - & Calculates the partitions of the data structures\\
\hline
HzCloudlet & Cloudlet & * Extends Cloudlet\\
\hline
HzVm & Vm & * Extends Vm\\
\hline
HazelSim & - & * Singleton of Hazelcast integration\\
\hline
HzObjectCollection & - & * Provides unified access to distributed objects\\
\hline
\end{tabular}
\end{table}

Hazelcast monitoring and heart beats are run on a separate thread, hence not interfering with the main thread that runs the simulations. Simulation objects, cloudlets and VMs were ported from Java lists to Hazelcast distributed maps. This enabled storing these objects in a distributed shared memory provided by Hazelcast spanning across the cluster. Instances of Hazelcast $IMap$ are used as the data structure. The core CloudSim class, $CloudSim$ is extended as $HzCloudSim$ to address the Hazelcast specific initializations. Similarly, $Cloudlet$ and $Vm$ are extended as $HzCloudlet$ and $HzVm$ respectively. This extended class hierarchy enabled modifying the internals of Vm and Cloudlet classes by sub-classing them to use Hazelcast distributed maps as the storage data structure, instead of Java lists, with instances of Hazelcast distributed executor service for distributed invocations of the methods. 

\subsubsection{Major Cloud Simulation Components}
Application layer provides sample cloud and MapReduce simulations, and structures that can assist developing further simulations on top of them. Existing CloudSim samples and applications can be ported to $Cloud^{2}Sim$ using this. 

\paragraph*{Compatibility Layer:}
A new package named ``compatibility'' composed of the core classes such as $HazelSim$ is placed inside CloudSim to integrate Hazelcast, Infinispan, and other new dependencies, and to enable multiple modes of operation (Such as Hazelcast or Infinispan based and regular CloudSim simulations). $HazelSim$ is the single class that is responsible for initiating the Hazelcast clusters and ensuring that the minimum number of instances are present in the cluster before the simulation begins. Hazelcast can also be configured programmatically for $Cloud^{2}Sim$ using $HazelSim$. $HzObjectCollection$ provides access to the distributed objects such as Hazelcast maps. $InfiniSim$ provides similar functionality for the Infinispan based distribution. The configuration file, $cloud2sim.properties$ is used to input MapReduce and CloudSim specific parameters such as the number of resources and users to be present in the simulation, such that simulations can be run with varying loads and scenarios, without need for recompiling.

\paragraph*{$Cloud^{2}Sim$ Core:}
The packages $cloudsim.hazelcast$ and $cloudsim.infinispan$ respectively integrate Hazelcast and Infinispan into the simulator. The concurrency layer consists of callables and runnables for asynchronous invocations to concurrently execute. As complex objects should be serialized before sending them to other instances over the wire, custom serializers are needed for $Vm$, $Cloudlet$, $Host$, $Datacenter$, and the other distributed objects to be able to distribute them across the instances, store and access them remotely in a binary format, effectively. The utilities module provides the utility methods used throughout $Cloud^{2}Sim$.

\paragraph*{Scheduling:}
The $scheduling$ package provides enhancements to the existing application scheduling capabilities of CloudSim. Matchmaking-based scheduling algorithms have to search through the complete object space to find a matching resource for the application requirements~\cite{mm,mm2}. The scheduling package handles scheduling in similar complex scenarios that involve searching large maps consisting of VMs, cloudlets, and the user requirements. Distributed application scheduling is done by the extended data center brokers that are capable of submitting the tasks and resources in a distributed manner. Moreover, strict matchmaking based algorithms and partial utility algorithms with matchmaking, require calculations and search for the objects, independent for each cloudlet. These are handled by the extended brokers.

DatacenterBroker and Datacenter are extended to provide a distributed execution. Extended brokers and their interaction with the resources and cloudlets are depicted in Figure~\ref{fig:classDiagram}.

\begin{figure}[!htbp]
\begin{center}
 \resizebox{0.7\columnwidth}{!}{
  \includegraphics[width=0.7\textwidth]{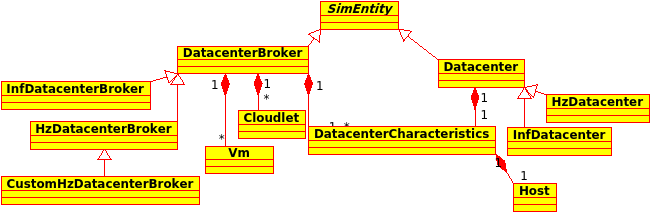}
 }
\end{center}
 \caption{Class Diagram of $Cloud^{2}Sim$ Brokers}
 \label{fig:classDiagram}
\end{figure}

\subsubsection{Distributed Execution of a Typical CloudSim Simulation}
The execution flow of a distributed simulation of an application scheduling scenario with Round Robin algorithm is shown by Figure~\ref{fig:flowmin}. It shows the simulation utilizing the core modules of $Cloud^{2}Sim$, and CloudSim to execute in a distributed manner. 
\begin{figure}[!h]
\begin{center}
 \resizebox{0.95\columnwidth}{!}{
  \includegraphics[width=0.95\textwidth]{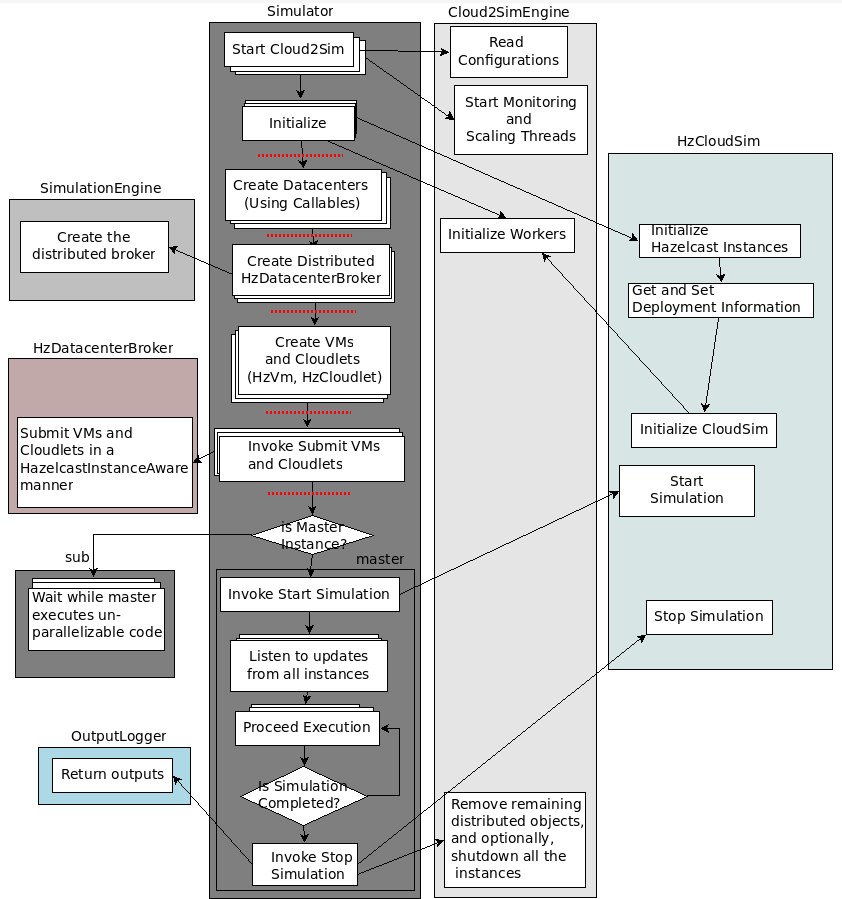}
 }
\end{center}
 \caption{Higher Level Execution flow of an application scheduler simulation}
 \label{fig:flowmin}
\end{figure}

A CloudSim simulation is started in $Cloud^{2}Sim$ by starting and initializing all the instances. Datacenters and distributed brokers are created concurrently. VMs and cloudlets are created in a distributed manner, from all the instances, where each instance holds a partition of entire VMs and cloudlets in it. Related entries are stored in the same instance or partition to minimize remote invocations. Created VMs and cloudlets are submitted by all the instances.

In a simulation such as Matchmaking based application scheduling~\cite{mm,mm2}, the major matchmaking phase consisting of matching the resources to the application can be entirely done in a distributed and independent manner. This is done by the broker in a distributed and partition-aware manner. This is the major workload of the simulation, as the simulation has to search through a large object space to find a match, where a distributed and concurrent execution helps. The searching and matching is done in each instance by the broker in the respective partitioned distributed object space.

Broker finally starts the core simulation. Since the VMs are distributed to multiple instances in Hazelcast distributed storage, the execution is distributed. However, the final outcome is presented by the master instance, collecting the outcomes from all the instances.

In a simulation where multiple VMs and cloudlets are created, and a cloudlet workload such as a matchmaking with a respective VM is involved, percentage of the independent cloudlet execution is very high among the entire execution, such that a distributed execution will provide much faster simulations.

\subsection{MapReduce Layer}
As Hazelcast MapReduce implementation is relatively new, the MapReduce layer has two implementation with Hazelcast and Infinispan, following the same design. It provides the MapReduce representation and implementations based on Hazelcast and Infinispan MapReduce modules. The minimal architecture of the MapReduce simulator of $Cloud^{2}Sim$ is depicted by Figure~\ref{fig:mapreduceImpl}, striping off the cloud simulation components. MapReduce Simulator can be configured with health monitoring and scaling. Hence, the execution time for varying the number of map() and reduce() invocations for various scenarios and simulations, as well as the health parameters such as load average and CPU utilization can be measured.
\begin{figure}[!htbp]
\begin{center}
 \resizebox{0.4\columnwidth}{!}{
  \includegraphics[width=0.4\textwidth]{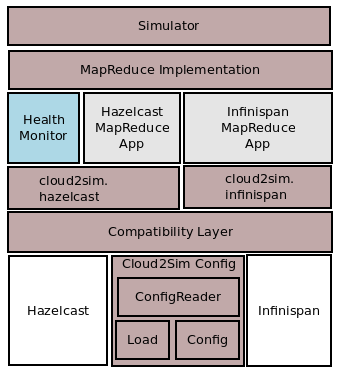}
 }
\end{center}
 \caption{Architecture of the MapReduce Component}
 \label{fig:mapreduceImpl}
\end{figure}

Both Hazelcast and Infinispan based MapReduce implementations have an $Initiator$ class that starts an instance of Hazelcast or Infinispan respectively and joins the main simulation cluster. While the $HzMapReduceSimulator$ or $InfMapReduceSimulator$ that runs from the master node coordinates and initiates the MapReduce jobs, the instances running $Initiator$ join the cluster and do the equal share of the jobs. The master node hosts the supervisor of the MapReduce job. The MapReduce implementation functions in verbose and non-verbose mode. In verbose mode, local progress of the individual map/reduce executions can be viewed from all the instances, where the final outcome is printed only to the master instance. Design of the $Cloud^{2}Sim$ MapReduce simulator and how it is integrated into $Cloud^{2}Sim$ is shown by Figure~\ref{fig:mrclass}.
\begin{figure}[!htbp]
\begin{center}
 \resizebox{0.60\columnwidth}{!}{
  \includegraphics[width=0.60\textwidth]{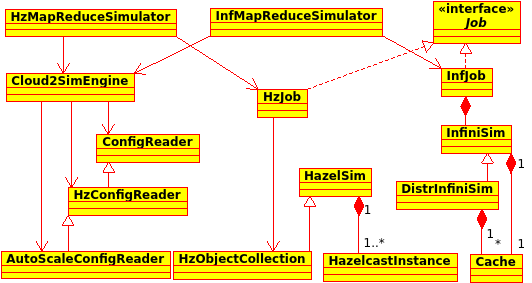}
 }
\end{center}
 \caption{Class Diagram of the MapReduce Simulator}
 \label{fig:mrclass}
\end{figure}

The default application used to demonstrate the MapReduce simulations is a simple word count application, which lets the user visualize different MapReduce scenarios. This default implementation can be replaced by custom MapReduce implementations. Figure~\ref{fig:mrhi} shows the alternatives and execution flow of the scalable MapReduce simulation platform.
\begin{figure}[!htbp]
\begin{center}
 \resizebox{0.45\columnwidth}{!}{
  \includegraphics[width=0.45\textwidth]{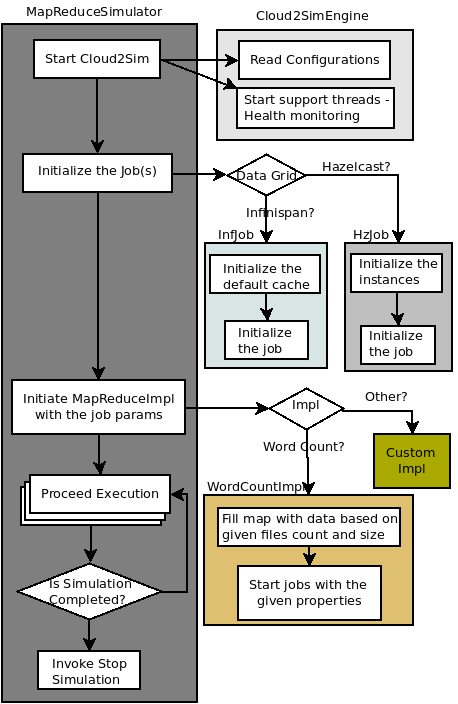}
 }
\end{center}
 \caption{Execution and Implementation Alternatives of the MapReduce Platform}
 \label{fig:mrhi}
\end{figure}

\subsection{Elasticity and Dynamic Scaling}
$Cloud^{2}Sim$ achieves elasticity through its implementations of dynamic scaling. Auto scaling and adaptive scaling are implemented by the packages $scale.auto$ and $scale.adaptive$. To prevent loss of information when the cluster scales in, synchronous backups are enabled by marking synchronous backup count as 1 in $hazelcast.xml$. This makes $Cloud^{2}Sim$ able to tolerate crashes, and avoid wasted work in long simulations, due to the availability of backups in different Hazelcast instances. Hazelcast stores the backups in different physical machines, whenever available, to minimize the possibility of losing all the backups during a hardware failure of a computer node. 

Since the time the instances are up and running can be measured at the individual node level, the cost of the simulation can be estimated, by assigning some cost values to the computing resources provided by the cluster instances. This is essentially viewing the service provided by the other nodes with the Hazelcast based IntelligentAdaptiveScaler as a cloud middleware Platform-as-a-Service. Moreover, the adaptive scaler design suits for any application, not limited to simulations. Hence this can be extended to use on any application that has a scaling requirement.

Distributed objects are removed by the user simulations as appropriate at the end of simulations. Hence it was decided not to use eviction policies of Hazelcast in $Cloud^{2}Sim$ by default, as it interferes with the user preference on dealing with the life-time of objects. Mostly the simulators clean the objects in the local instance and also shut down the local Hazelcast instance. The distributed objects still remaining in the Initiator instances after the simulation, are removed. This enables the Initiator instances to join the other simulations without the need to restart them.

\paragraph*{Summary:}
Cloud and MapReduce simulations can be executed on top of in-memory data grids, that executes over the computer clusters. Cycle sharing of the instances in the cluster, inspired by volunteer computing, is used as the model to achieve a scalable, adaptive, and elastic middleware platform for the simulations. Hazelcast and Infinispan are integrated into core CloudSim as a compatibility layer for a seamless integration and invocation of cloud simulations.

Multiple simulations can be executed in parallel using the $IntelligentAdaptiveScaler$ approach, where a single $Coordinator$ can ensure each tenant, that is represented by a cluster, has adequate resources allocated for an optimal execution. Multi-tenanted deployments of $Cloud^{2}Sim$ enable multiple distinct cloud and MapReduce simulations to be coordinated and scaled from a single health monitor and adaptive scaler, where a global view of the deployment is available to the $Coordinator$ node, as it consists of instances in all the clusters that it coordinates.
\chapter{Implementation}
\label{chap:imp}

Based on the design, $Cloud^{2}Sim$ was implemented as a concurrent and distributed cloud and MapReduce simulator. Section~\ref{sec:4cd} looks into the implementation details of the concurrent and distributed cloud simulator more thoroughly. Scalability and elasticity are important aspects in a distributed execution, so that the system will scale seamlessly, when more instances are introduced. The MapReduce simulator has two implementations based on Hazelcast and Infinispan, which are built independent from the cloud simulations. Both Hazelcast and Infinispan based implementations are independent from each other, letting the users pick the one that better suits their requirements. Since the same simulation code will run in both implementations, this also provides an opportunity to compare the efficiency of Hazelcast and Infinispan to use as MapReduce platforms, as the design and tasks are identical for both implementations. Section~\ref{sec:4ms} looks into the MapReduce implementations, integrations, and simulations. 

Section~\ref{sec:4emp} discusses how scalability and elasticity are achieved in the middleware platform. The IntelligentAdaptiveScaler is an architecture and solution that we have developed to provide an adaptive scaling strategy, which is used to scale $Cloud^{2}Sim$ across the available nodes in the cluster. Though this is used for cloud simulations and MapReduce executions of $Cloud^{2}Sim$, this can be adopted to use in any applications. The implementation and integration of the adaptive scaling strategy and solution, are presented further.

  %
  %

\section{Concurrent and Distributed Cloud Simulator}
\label{sec:4cd}
CloudSim simulations can run on $Cloud^{2}Sim$ with minor changes to facilitate distribution. Distributing the simulation environment has been implemented using an incremental approach. The CloudSim trunk version was forked and used in the implementation. A complete distributed cloud simulator was built with Hazelcast, having CloudSim as the core simulation module.

Hazelcast version 3.2 and Infinispan version 6.0.2 were used in the implementations and evaluations. JGroups is a reliable multicasting toolkit~\cite{ban2002jgroups} that is used internally by Infinispan for clustering and grouping, and version 3.4.1 is used by Infinispan 6.0.2. CloudSim Loggers are used in logging, throughout $Cloud^{2}Sim$ project, to log the outputs. Built using Apache Maven, the project is hosted on SourceForge\footnote{Checkout the source code at \url{https://sourceforge.net/p/cloud2sim/code/ci/master/tree/}, with user name, ``cloud2sim'' and password, ``Cloud2Simtest''.}, with the Git distributed version control system.

\subsection{Concurrent and Distributed Storage and Execution}
Though the data center, VM, and cloudlet creation as well as the scheduling can be done concurrently in a multi-threaded environment, CloudSim does not exploit this as it provides a sequential execution for most of these workflows. $Cloud^{2}Sim$ exploits the multicore systems by using multiple threads to handle these concurrently, while having respective concurrent barriers to ensure accurate simulation outcomes.

Initially, sample concurrent simulations were implemented, with concurrent data center creation, concurrent initialization of VMs and cloudlets, and submission of them to the brokers. Classes extending the $Runnable$ and $Callable$ interfaces were used to submit the VMs and cloudlets concurrently. Though the initialization of threads and executor frameworks introduced an overhead for small simulations, it provided a speed-up for the larger simulations. Very small simulations do not require distributed execution, as they perform reasonably well, and were never the target of this work. Simulations that fail to execute or perform poorly due to the processing power requirements on a single thread, perform much better on the concurrent environments utilized by $Cloud^{2}Sim$. Hence, the overheads imposed by the initializations is not a limitation to usability, as the performance gain is higher. Sample prototype developments with concurrent creation of data centers showed an increased performance, overcoming the overheads.

Hazelcast was initially just used to provide a distributed storage, using one instance of $Simulator$ and multiple instances of $Initiator$, following the \textit{Simulator - Initiator based Strategy}, as described in Section~\ref{ssec:3partition}. However, distributing the complex VM and cloudlet objects introduced communication and serialization costs for most of the CloudSim simulations, though it fit the MapReduce simulations. Hence, a $SimulatorSub$ was implemented for each $Simulator$, where the execution that can be distributed rather was started from multiple instances. Finally, \textit{multiple Simulator instances strategy} was implemented where the first instance to join the cluster becomes the master and executes the core fractions of the logic which must not be distributed, decentralized, or run in parallel for a correct execution of the simulation. This strategy fits the CloudSim simulations, as there is no requirements to have a specific instance to be statically allocated as the master instance.

Hazelcast and Infinispan based clustering is done via TCP and UDP multicasting. When groups are formed inside a single node, UDP multicasting is used for detecting the instances and creating the cluster. Properties of the Hazelcast cluster such as whether the caching should be enabled in the simulation environment, and when the unused objects should be evicted from the instances are configured by $hazelcast.xml$. Similarly, $infinispan.xml$ is used to configure the Infinispan cache. When running across the nodes in a physical cluster, TCP based discovery is used, where the instances are predefined in hazelcast.xml for Hazelcast based implementation, and in jgroups-tcp-config.xml for Infinispan based implementation. JGroups~\cite{ban2002jgroups} is configured as the core group communication technology beneath Infinispan clustering, and the respective TCP or UDP configuration files are pointed from infinispan.xml.

Hazelcast $IExecutorService$ was utilized to make the execution distributed. While MapReduce executions were effectively executed in the \textit{Simulator - Initiator based strategy}, cloud simulations rather follow a model where all the instances initiate and send logic fractions. Initially implemented as different classes, following \textit{Simulator - SimulatorSub based strategy}, the master and other instances were later unified, following the \textit{multiple Simulator instances Strategy}, such that a same $Simulator$ class can be run from all the instances.

Callables and runnables were made to implement $HazelcastInstanceAware$ interface, to ensure the members of the clusters executed part of the logic on the data partition that is stored in themselves, to minimize remote invocation, by increasing data locality. In a distributed environment, near-caching might improve the performance by caching the frequently accessed objects in the object space provided by the same instance or node. However, near-caching is disabled in $Cloud^{2}Sim$ when running in multiple nodes, to avoid the increased memory consumption and to avoid reading the inconsistent outdated objects. Instead of using near-caching, $Cloud^{2}Sim$ optimizes the data locality of the distributed objects by storing the related objects together, as they frequently access each other. 

\subsection{Serialization}
$Cloud^{2}Sim$ uses $BINARY$ in-memory format for cloud simulations as $Cloud^{2}Sim$ contains custom objects that are mandatory to be stored serialized. Since the MapReduce simulator does not use complex objects that are not serializable, Hazelcast is configured with $OBJECT$ in-memory format for MapReduce simulations. This eliminates most serialization costs from the MapReduce executions. As the CloudSim objects to be distributed are custom objects that cannot be directly serialized, custom serializers were written for them, extending the Hazelcast StreamSerializer interface. Custom serializers are registered with the respective classes that they serialize using hazelcast.xml, in the $<serialization>$ section, as shown below.

{\fontsize{10}{10}\selectfont
\begin{lstlisting}
<serializers>
  <serializer type-class="pt.inesc_id.gsd.cloud2sim.hazelcast.HzVm">
    pt.inesc_id.gsd.cloud2sim.serializer.VmXmlSerializer
  </serializer>
  <serializer type-class="org.cloudbus.cloudsim.Host">
    pt.inesc_id.gsd.cloud2sim.serializer.HostXmlSerializer
  </serializer>
  <serializer type-class="org.cloudbus.cloudsim.CloudletScheduler">
    pt.inesc_id.gsd.cloud2sim.serializer.CloudletSchedulerXmlSerializer
  </serializer>
  <serializer type-class="pt.inesc_id.gsd.cloud2sim.hazelcast.HzCloudlet">
    pt.inesc_id.gsd.cloud2sim.serializer.CloudletXmlSerializer
  </serializer>
  <serializer type-class="org.cloudbus.cloudsim.UtilizationModel">
    pt.inesc_id.gsd.cloud2sim.serializer.UtilizationModelXmlSerializer
  </serializer>
  <serializer type-class="org.cloudbus.cloudsim.Datacenter">
    pt.inesc_id.gsd.cloud2sim.serializer.DatacenterXmlSerializer
  </serializer>
</serializers>
\end{lstlisting}
}

Classes in the serializer package contains the relevant properties of the constructor of the class that implements the respective data structure, as shown below.

{\fontsize{10}{10}\selectfont
\begin{lstlisting}
String[] propertyNames = new String[] { "id", "userId", "mips", 
	"numberOfPes", "ram", "bw", "size", "vmm", "cloudletScheduler" };
encoder.setPersistenceDelegate(HzVm.class, 
	new DefaultPersistenceDelegate(propertyNames));
\end{lstlisting}
}
\subsection{Partitioning Approaches}
As discussed in the previous chapter, different instances function as the master instance and the slave instances in the simulation task, based on the order that they were started and joined the cluster. The first to join the cluster becomes the master, where the others become the slaves, or subs.

Partitioning of data and execution is calculated iteratively for each instance. The number of instances currently in the cluster is tracked by an instance of distributed map, called $deploymentList$. An instance will have an offset value assigned to it, which is the number of instances that have joined previously. Hence the offset of the first instance will be zero and initial ID of the partition will be zero as well. Final ID of the data partition of the instance that joins last, will be same as the last ID of the distributed data structure.

{\fontsize{10}{10}\selectfont
\begin{lstlisting}
    /**
     * Gets the initial value of the partition
     *
     * @param noOfParams, total number of entities in the specific parameter.
     * @param offset, the offset
     * @return the initial value of the partition
     */
    public static int getPartitionInit(int noOfParams, int offset) {
        return (int) (offset * Math.ceil((noOfParams /
                (double) HzConstants.NO_OF_PARALLEL_EXECUTIONS)));
    }

    /**
     * Gets the final value of the partition
     *
     * @param noOfParams, total number of entities in the specific parameter.
     * @param offset, the offset
     * @return the final value of the partition
     */
    public static int getPartitionFinal(int noOfParams, int offset) {
        int temp = (int) ((offset + 1) * Math.ceil((noOfParams /
                (double) HzConstants.NO_OF_PARALLEL_EXECUTIONS)));
        return temp < noOfParams ? temp : noOfParams;
    }
\end{lstlisting}
}
The partition logic allows dynamic scaling, where the instances can join and leave during execution. This implementation enables easy integration of auto scaling implementations into the simulation. By default, the back up count is set to zero. It is set to 1 when the dynamic scaling is enabled, to avoid data loss due to scale ins.

\subsection{Execution Flow}
$Cloud2SimEngine$ is started as the initial step of $Cloud^{2}Sim$ cloud simulations. $Cloud2SimEngine.start()$ starts the timer and calls $HzConfigReader$ to read the configurations. If health checks are enabled, it starts the health monitor thread, to periodically monitor the instance status and report as configured. If adaptive scaling is enabled, it also starts the AdaptiveScalerProbe in a separate thread, to communicate with the $IntelligentAdaptiveScaler$ instances in the other nodes to adaptively scale the simulation. It finally initializes HzCloudSim, where Hazelcast simulation cluster is initialized with the simulation job, and CloudSim simulation is started.

Data centers are created concurrently. Brokers extending $HzDatacenterBroker$ create instances of $HzVm$ and $HzCloudlet$ and start scheduling in a distributed manner, using all the instances in the simulation cluster. The core simulation is started using $HzCloudSim.startSimulation()$, and executed by the master instance. When the simulation finishes, the final output is logged by the master instance. Based on the simulation, the instances are either terminated or their distributed objects are cleared and the instances are reset for the next simulation. Figure~\ref{fig:flow} depicts the execution flow of the application scheduling, along with the respective methods, extending the higher level execution flow shown in Figure~\ref{fig:flowmin}.

\begin{figure}[!htbp]
\begin{center}
 \resizebox{0.85\columnwidth}{!}{
  \includegraphics[width=0.85\textwidth]{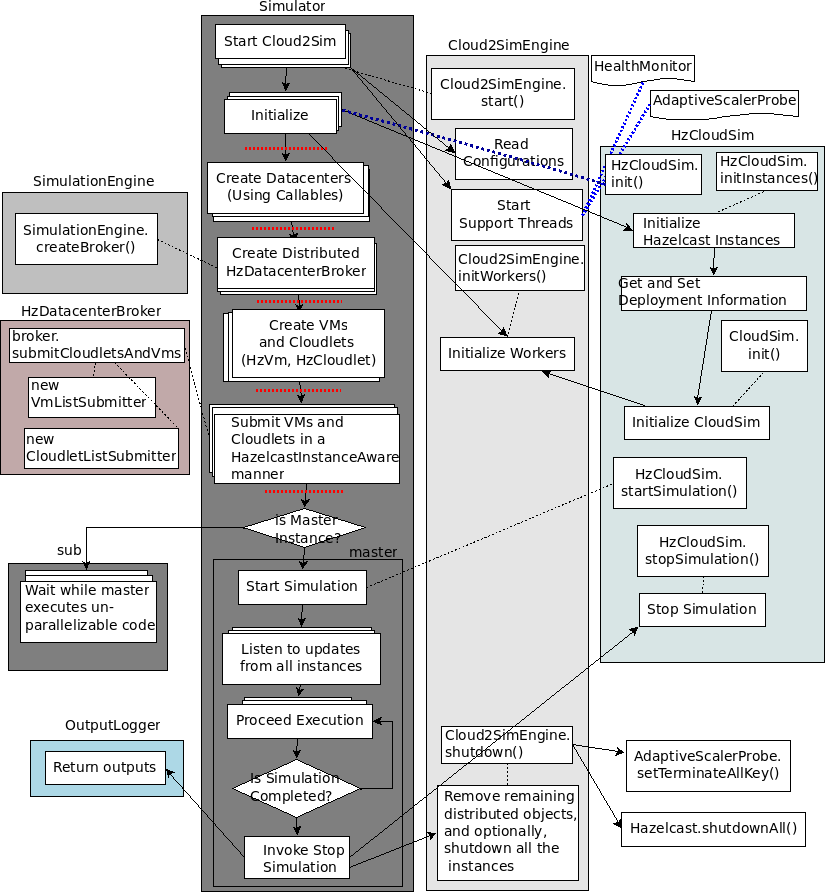}
 }
\end{center}
 \caption{Execution flow of an application scheduler simulation with the Implementation}
 \label{fig:flow}
\end{figure}

\paragraph*{Trade-offs Analysis}
Initial challenges involve the overheads imposed by the distributed execution such as the serialization costs, and optimization measures were taken to minimize these. External dependencies were used unmodified, as a design decision, to increase the portability and to minimize the maintenance costs. Also, changes to CloudSim core was avoided, with only minimal changes to enable extending the data structures, such as marking the methods from private access to protected, and marking the final classes not final to be able to extend them to enable distributed executions.

Custom serializers were implemented and registered with the classes of the respective objects that they intend to serialize. Partition-aware thread invocations such as $IExecutorService.executeOnKeyOwner()$ execute the operation on the instance that holds the distributed object, instead of accessing it remotely on a locally created copy of the object, as it is the default behaviour when invoking logic on data partitions without delegating the invocation to an executor service. This reduces the communication cost. Due to these trade-offs, simulations having a huge memory or processor overhead to run on a single node exhibit positive scalability with faster executions on multiple instances, where small simulations that are fast enough running on single instances naturally slow down on distributed executions.

\subsection{Distributing Custom CloudSim Simulations}
More complex CloudSim simulations have been developed by the researchers to prototype their work, with custom developments and extended brokers. Power-aware simulations such as Dynamic Voltage Frequency Scaling (Dvfs) use CloudSim $PowerDatacenterBroker$ as the broker, extending $DatacenterBroker$. $PowerDatacenterBroker$ overrides $processVmCreate()$ method of the broker. 

More brokers are developed for complex simulations. To make a distributed execution with Hazelcast-based $Cloud^{2}Sim$, brokers should be developed extending $HzDatacenterBroker$ instead of $DatacenterBroker$. Similarly, $HzCloudSim$ represents the core $Cloud^{2}Sim$ class that extends $CloudSim$ class, which should be started for the simulation execution.
  %
  %
\section{MapReduce Simulator}
\label{sec:4ms}
MapReduce Simulator has two different implementations, based on Hazelcast and Infinispan. A basic MapReduce word count application was implemented using Hazelcast and Infinispan and incorporated into $Cloud^{2}Sim$. Complex MapReduce scenarios were simulated using this small application. 

\subsection{Infinispan Initialization}
Hazelcast integration is already discussed with the cloud simulations, as it is common for both cloud and MapReduce simulations. Infinispan is integrated using the compatibility layer in CloudSim, to facilitate later migration of $Cloud^{2}Sim$ to Infinispan. This also enables the same design and architecture for both Hazelcast and Infinispan based distributions.

$Infinisim$ in the compatibility layer configures the $DefaultCacheManager$ of Infinispan, using the $infinispan.xml$ pointed by $cloudsim.properties$. A transactional cache is created from the cache manager. An instance of cache in Infinispan is similar to an instance in Hazelcast. Multiple instances of Cache form a cluster and execute the jobs. Simulator and Initiator instances are created using the same configurations. The cache instance initialized by the master node acts as the supervisor of the MapReduce jobs, and distributes the tasks across the Initiator instances.

\subsection{MapReduce Implementation}
InfJob and HzJob classes implement the Job interface. HzJob and InfJob get the job information, and the real implementation is done by the classes in mapreduce.hazelcast.impl and mapreduce.infinispan.impl packages. Due to this separation, MapReduce implementation can be customized to another implementation by the users instead of the current word-count based implementation, if they prefer. 

The default word count MapReduce application reads and counts big files that are stored in the folder pointed by the property $loadFolder$ in $cloud2sim.properties$. This default implementation can be changed with different sample applications. Huge text files such as the files collected from USENET Corpus were used~\cite{usenet}. Current implementation stands as a decent working sample, following the CloudSim approach of providing examples. 

\subsection{Customizing the MapReduce job invocations}
MapReduce size refers to the number of lines from the load files that are read for the MapReduce tasks. The simulator reads up to a specific line in each of the files, as indicated in the $cloud2sim.properties$. In a simple MapReduce simulation, map() is invoked multiple times as many as the files present in the load folder, or a smaller number defined by the user. Invocations of map() increase with the number of files present in the load folder, which is pointed by the loadFolder property in the configuration file. 

Invocations of reduce() increases with the number of lines or the size of the task specified by the user. By using duplicate files, invocations of map() are increased, keeping the reduce() invocations constant. Increasing the number of lines read from the same set of files increases the reduce() invocations, keeping the map() invocations constant. By increasing both the size and number of files, both map() and reduce() invocations can be increased simultaneously. Thus, effect of changing these method invocations and parameter spaces on performance as well as simplicity can be monitored.

This simple configuration helps develop a sample MapReduce application with varying number of map() and reduce() invocations. More complex MapReduce applications can be visualized by this simulator, which is a simple MapReduce application implemented on Hazelcast and Infinispan. Scalability by increasing the number of physical nodes and its effects on the execution time and status change of the instances such as load average can be monitored.

The execution flow of a MapReduce simulation using Infinispan implementation is shown by Figure~\ref{fig:mrflow}. Hazelcast execution is similar, with the respective classes of Hazelcast implementation.
\begin{figure}[!htbp]
\begin{center}
 \resizebox{0.7\columnwidth}{!}{
  \includegraphics[width=0.7\textwidth]{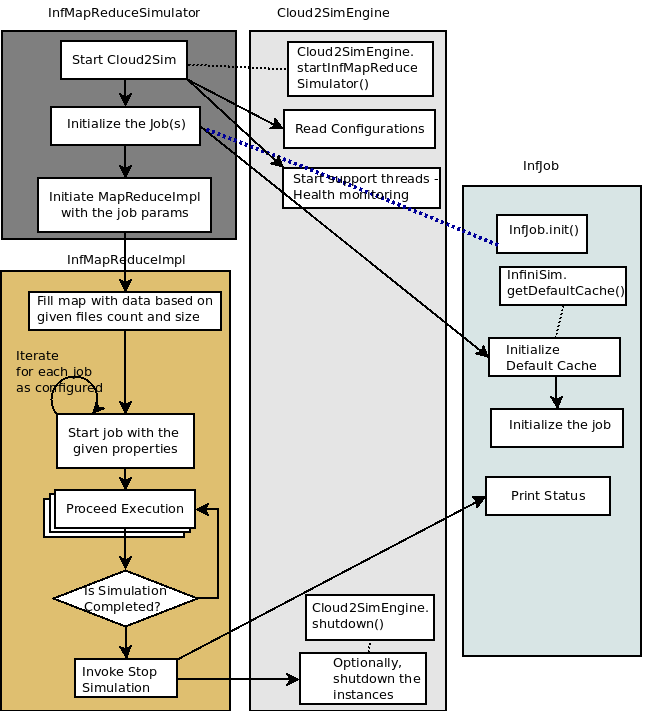}
 }
\end{center}
 \caption{Execution flow of a MapReduce simulation with the Infinispan Implementation}
 \label{fig:mrflow}
\end{figure}

\section{Elastic Middleware Platform}
\label{sec:4emp}
Scalability is a major requirement for the simulator for an effective execution of the distributed simulations. The in-memory data grid platforms, Hazelcast and Infinispan, are integrated such that they offer resources from the nodes that they run on, to contribute to the simulation deployment. Larger simulations are executed with more instances started when starting the simulations. The simulations scale out to more instances as more and more instances are started manually, or by the dynamic scaling configured with the health monitor. 

The elasticity of a platform is defined as the ability of the platform to scale dynamically, along with the increasing or decreasing resource requirements. An efficient elastic solution provides a cost-effective strategy for cloud deployments. Moreover, elasticity optimizes the resource allocation to the simulations in the simulator platform, as the resources are registered with the simulator according to the execution requirements. We will look at the implementation details of the elastic middleware platform and how elasticity is achieved by the developed middleware platform to cater for the dynamically changing load. The developed middleware platform and elastic strategy is generic enough such that it is not limited to CloudSim simulations, but also can be used to scale out the other applications. This work integrates the elasticity to cloud and MapReduce simulations.

\subsection{Health Monitoring}
Hazelcast contains a health monitoring module itself, which periodically logs heart beat information. This module requires extensions to be able to be used for $Cloud^{2}Sim$ scaling requirements. Hence, a health monitoring module using $com.sun.management.OperatingSystemMXBean$ was developed, following the Java monitoring and management extensions. The health monitor module provides implementation for a few parameters such as CPU load, and also provides an API to extend it further. If process CPU load is used as the measure for scaling, health monitor will notify the dynamic scaling modules, when the process CPU load exceeds the maximum defined in $cloud2sim.properties$ file. Similarly, it will also notify when it goes below the minimum defined. This can also be done using the other system characteristics monitored by the health monitor, based on the scaling policies and scaling modes. 

This exhibits an adaptive behavior, as once a new instance is spawned, it will wait for a user-defined period which is usually longer than the time interval for health checks, for the next scaling action. This longer wait between scaling decisions prevents cascaded scaling and jitter where multiple instances are added or removed at once, or within a very short period of time interval, during the time taken for the effect of the change of the instance count to be reflected. 

The gap between the high and low thresholds are kept reasonably high, to prevent the jitter effect, where instances are added and removed frequently, as the high and low thresholds are frequently met. The health monitor configuration provides means to configure the scaling and monitoring to fit the application requirements and extend the module further to fine tune according to the application requirements.

\subsection{Adaptive Scaling}
Scaling decisions should be atomic, executing in the distributed environment, to make the adaptive scaling work without simultaneously starting or shutting down multiple instances. An instance of Hazelcast $IAtomicLong$, a distributed implementation of the Java concurrent atomic object, is used as the flag to get and set the scaling decisions, to scale-out or to scale-in, as shown in Figure~\ref{fig:subcluster}. Other concurrent atomic objects such as $AtomicBoolean$ are not available as a distributed object yet, though there is a feature request to $IAtomicBoolean$ in Hazelcast community. 

\begin{figure}[!htbp]
\begin{center}
 \resizebox{0.5\columnwidth}{!}{
  \includegraphics[width=0.5\textwidth]{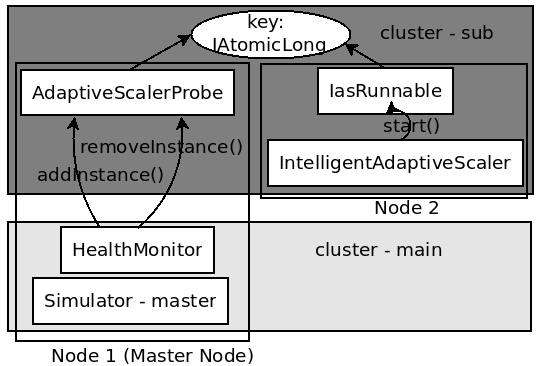}
 }
\end{center}
 \caption{Adaptive Scaling Modules}
 \label{fig:subcluster}
\end{figure}

The same object is used as the flag for scaling-in and scaling-out decisions, as the scaling in and scaling out scenarios do not occur at the same time. The AdaptiveScalerProbe thread sets the value of this object to -1 or 1, for scaling-in or scaling-out request respectively. This object is set back to 0 by the IntelligentAdaptiveScaler of one of the instances, when it was noticed and the scaling action is initiated. Upon the completion of the simulation, AdaptiveScalerProbe of the master instance sets this flag to the $TERMINATE\_ALL\_FLAG$ value (by default, -999), notifying the other instances in the main cluster to shutdown, while the last instance to shut down clears up the distributed objects in the sub-cluster. Non-atomic distributed objects were used for other scaling decisions, to avoid slowing down the scaling process with locks.

\subsection{Review on Major Implementation Challenges}
During the attempt of implementing group migration of the Hazelcast instances, Hazelcast exhibited a Split-brain syndrome, where two sub-clusters emerged from a single cluster and existed without the knowledge of the instances in the other sub-cluster. However, this did not harm the simulation execution as distributed objects have synchronous backup in the elastic mode. After a short period, the sub-clusters merged themselves, recovering from the split. This behavior was not observed in the implementation of IntelligentAdaptiveScaler approach, which is the chosen approach for the implementation.

Hazelcast Management Center tool was often interrupting the sleep() in the threads of the IntelligentAdaptiveScaler. It also caused memory leaks when running with multiple clusters. Hence, the Management Center tool is disabled during the final experiments involving scalability.

\paragraph*{Summary:}
$Cloud^{2}Sim$ is implemented as a distributed and concurrent cloud and MapReduce simulator. MapReduce executions are inherently distributable. Cloud simulations are distributed in a way that their accuracy is not affected, while maximizing the fractions of the execution that can be distributed. Tightly coupled core simulation fragments are not distributed, as the resultant coordination and communication costs will be dominant. Cloudlet workloads and Datacenter brokers can be distributed very effectively.

The Hazelcast based distributed cloud simulator is implemented along with prototype deployments and samples. Infinispan is integrated into CloudSim, such that it can be used to implement the middleware platform to scale the simulator, exploiting the same design. Two implementations of MapReduce simulators exist with Hazelcast and Infinispan MapReduce implementations as the core modules. The simulator platform is implemented as an elastic middleware platform for cloud and MapReduce simulations, but it can be extended for other applications as well.
\chapter{Evaluation}
\label{chap:eval}
We will discuss the experiments carried ahead to test the effectiveness of the solution in this chapter. Section~\ref{sec:5cs} discusses the CloudSim simulations. Based on the size of the simulations and load of the tasks, different patterns of scalability were observed. Section~\ref{sec:5ms} compares and evaluates the Hazelcast and Infinispan based MapReduce implementations, on their performance and scalability. Section~\ref{sec:5rw} compares $Cloud^{2}Sim$ features with the features implemented by related, similar and complimentary work.
  %
  %

\paragraph*{Deployment Environment:}
\label{sec:5de}
A computer cluster with 6 identical physical nodes (Intel(R) Core(TM) i7-2600K CPU @ 3.40GHz and 12 GB memory) was used for the evaluation. Multiple simulations were experimented on the system using 1 to 6 nodes. Each node executed one Hazelcast or Infinispan instance, except during the experiments involving multiple instances in a single node.
  %
  %

The master node always completes the last, as the final outcome is printed by the master node in the simulations considered. Time taken by the master node is noted down, as the other nodes finished the execution before the master. $nohup$ was used to start the process to avoid interrupts, and the output was directed to an output file called nohup.out, which was analysed upon completing the simulations or viewed interactively using tail. Accuracy of the simulations was confirmed through the outcomes - for example, the scheduling decisions for cloud simulations and the output of the MapReduce computation for the MapReduce simulations.

\begin{lstlisting}
 nohup java -classpath cloudsim-3.1-SNAPSHOT.jar:
    lib/hazelcast-3.2.jar:cloud2sim-1.0-SNAPSHOT.jar 
    pt.inesc_id.gsd.cloud2sim.mapreduce.hazelcast.HzMapReduceSimulator > nohup.out &

 tail -f nohup.out
\end{lstlisting}

\section{CloudSim Simulations}
\label{sec:5cs}
A round robin application scheduling simulation with 200 users and 15 data centers, was built and configured, with and without a cloudlet workload, for a varying number of VMs and cloudlets. It was initially evaluated on CloudSim, and then on $Cloud^{2}Sim$ with different number of nodes. 

Table~\ref{table:execTime} shows the time taken to simulate the scenario in CloudSim and $Cloud^{2}Sim$. CloudSim outperformed $Cloud^{2}Sim$ in the base execution without a workload, due to the dominant inherent coordination overload involved in $Cloud^{2}Sim$. $Cloud^{2}Sim$ with multiple nodes showed a considerable 10-fold improvement in the execution time when the cloudlets contained a relevant workload to be simulated once scheduled. Time taken (in seconds) for an experiment in $Cloud^{2}Sim$ with 1, 2, 3, and 6 nodes as well as in CloudSim is depicted by Table~\ref{table:execTime} for 200 VMs and 400 cloudlets.

\begin{table}[!ht]
\caption{Execution time (sec) for CloudSim Vs. $Cloud^{2}Sim$}
\label{table:execTime}
\begin{tabular}{|c||c| |c|}
\hline
Deployment & Simple Simulation & Simulation with a cloudlet workload\\
\hline
CloudSim & 3.678 & 1247.400\\
$Cloud^{2}Sim$ (1 node) & 20.914 & 1259.743 \\
$Cloud^{2}Sim$ (2 nodes) & 16.726 & 120.009 \\
$Cloud^{2}Sim$ (3 nodes) & 14.432 & 96.053 \\
$Cloud^{2}Sim$ (6 nodes) & 20.307 & 104.440 \\
\hline
\end{tabular}
\end{table}

In a simulation where each cloudlet does a complex job, the time taken for the simulation increases with the number of cloudlets. With the number of VMs fixed at 200, simulation time taken on 1 - 6 nodes was measured. Figure~\ref{fig:cloudlet} depicts how the application scales with varying number of cloudlets. 

\begin{figure}[ht]
\begin{center}
 \resizebox{0.6\columnwidth}{!}{
  \includegraphics[width=0.6\textwidth]{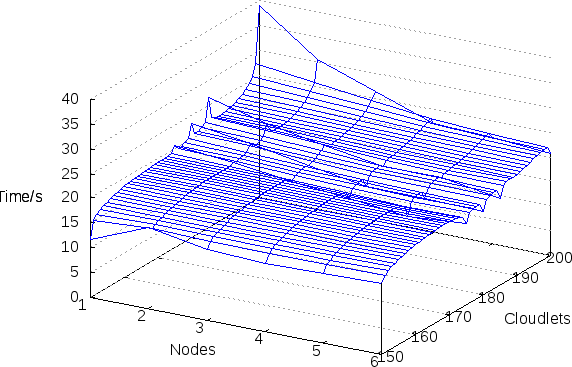}
 }
\end{center}

 \caption{Simulation of Application Scheduling Scenarios}
 \label{fig:cloudlet}
\end{figure}

When the number of cloudlets was 150, increasing the number of nodes to 2 increased the execution time, which later goes down when the nodes were further increased from 3 to 6. The initial negative scalability was due to the prominent costs such as serialization, communication, and coordination costs overshadowing the performance gain due to the distributed execution, as formulated in Section~\ref{sec:3perf}. When the size of the simulation was further increased up to 200 cloudlets, a perfect positive scalability was observed, as fraction of the code that can be distributed increases with the number of cloudlets with cloudlet workloads. As the size of the simulation is increased, performance is seen increasing with the number of nodes, depicting the suitability of the distributed execution model for larger simulations.

\subsection{Distributed Execution of Round Robin Application Scheduling}
The experiment was repeated with different combinations of VMs (from 100 - 200) and Cloudlets (from 100 - 400), with and without a complex mathematical operation to be performed for each cloudlet as a load (parameter `isLoaded' is set to true, for a cloudlet workload). Four distinct cases of scalability were noticed, as described below. 

\subsubsection{Success Case (Positive Scalability)}
Figure~\ref{fig:CASEa} depicts the scenarios of (noOfVMs = 200, noOfCloudlets = 400, isLoaded = true) and (noOfVMs = 100, noOfCloudlets = 200, isLoaded = true), where the time taken for simulation is decreasing with the number of nodes. This is a desired scenario of scaling where the task is so much CPU intensive for each cloudlet to handle in a single node, such that introducing more nodes distribute the tasks effectively, reducing the simulation time.

\paragraph*{\textbf{Dynamic Scaling}}
With the dynamic scaling enabled, this case introduced more instances into the execution, as the load goes high. Memory used by the application as a percentage of the total memory used, was used as the health monitoring measure. With the adaptive scaling, the environment of 200 VMs and 400 cloudlets with load, scaled up to 3 instances, for a CPU utilization of 0.20, even when more than 3 instances were included in the sub-cluster. Reducing the maximum threshold made the main-cluster to scale out earlier and faster, involving all the available instances to the simulation. Figure~\ref{fig:CASEa} shows the time taken for the simulations with and without adaptive scaling. As shown by Figure~\ref{fig:CASEa}, the execution time is converging as more nodes are added. Hence, introducing further nodes beyond a certain maximum number of nodes may not be economically feasible, and at a point this may become the case 3, which is explained below as the \textit{common case}.

\begin{figure}[ht]
\begin{center}
 \resizebox{0.5\columnwidth}{!}{
  \includegraphics[width=0.5\textwidth]{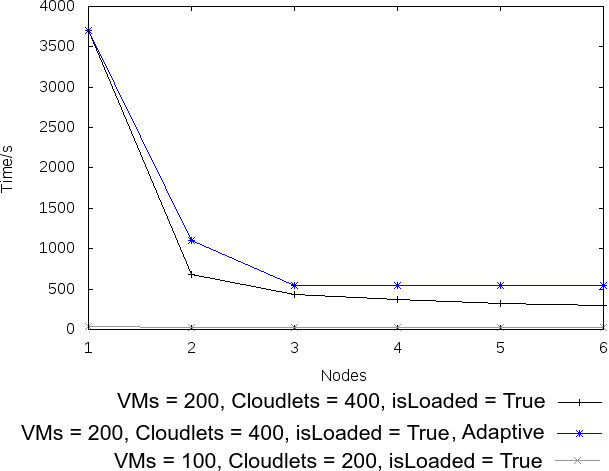}
 }
\end{center}

 \caption{Distributed Execution - Positive Scalability}
 \label{fig:CASEa}
\end{figure}

Adaptive scaling was not observed in the other cases, except when the maximum process CPU load is reduced below 0.15 from the configurations. This shows that a single instance was sufficient to run the sample simulations of the other 3 cases discussed below. For the scenario of scale ins, synchronous backups should be enabled to prevent the data loss, which eliminates the possibility of a fair comparison, as the simulations with the fixed number of instances are run with no backups. Hence, the low threshold was kept low enough during the experiment, such that there were no scale ins. Scale in was observed, when the minimum process CPU load was increased beyond 0.02.

\paragraph*{Load Average:}
With adaptive scaling configured, load average was logged during the execution. Table~\ref{table:loadavg} shows the load averages observed during and after the scaling events, for the simulation environment with 6 nodes available. Up to 3 nodes were involved in the simulation by the IntelligentAdaptiveScaler. Waiting time acts as a buffer to prevent cascaded scaling events. Health is monitored periodically, except during the buffer time introduced immediately following the scaling events. These intervals are configured to fit the requirements and the nature of the simulation.
\begin{table}[!t]
\caption{Load averages with Adaptive Scaling on 6 nodes}
\label{table:loadavg}

\begin{tabular}{|c| |c| |c| |c| |c|}
\hline
\textbf{Number of} & \textbf{Master } & \textbf{Spawned } & \textbf{Spawned } & \textbf{Event} \\
\textbf{Instances} & \textbf{ Instance - I0} & \textbf{ Instance - I1} & \textbf{ Instance - I2} & \textbf{} \\
\hline
\textbf{1} & 0.30 & - & - & Spawning Instance - I1\\
\hline
\textbf{2} & 0.30 & 0.24 & - & Waiting Time \\
\hline
\textbf{2} & 0.25 & 0.24 & - & Spawning Instance - I2 \\
\hline
\textbf{3} & 0.23 & 0.23 & 0.13 & Waiting Time \\
\hline
\textbf{3} & 0.21 & 0.19 & 0.13 & Health Monitoring \\
\hline
\textbf{3} & 0.09 & 0.18 & 0.09 & Health Monitoring \\
\hline
\textbf{3} & 0.06 & 0.18 & 0.08 & Health Monitoring \\
\hline
\end{tabular}
\end{table}

\subsubsection{Other Cases of Scalability}
Figure~\ref{fig:CASEb} depicts 3 distinct cases of scalability, where the execution time changes in different patterns with the increasing number of nodes. The scenarios are analyzed below.
\begin{figure}[ht]
\begin{center}
 \resizebox{0.5\columnwidth}{!}{
  \includegraphics[width=0.5\textwidth]{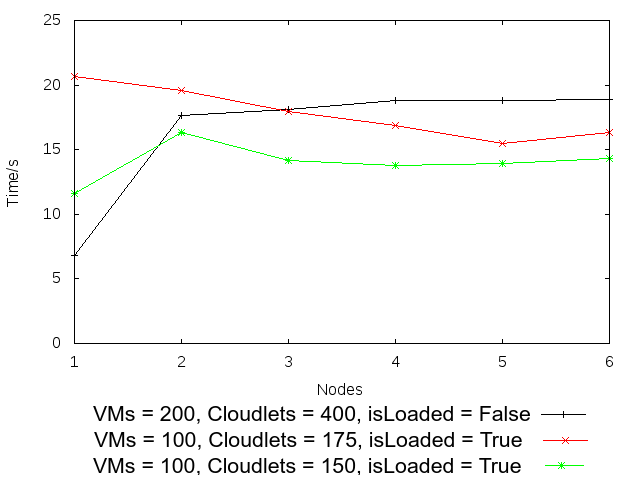}
 }
\end{center}

 \caption{Distributed Execution - Different Patterns of Scaling}
 \label{fig:CASEb}
\end{figure}

\paragraph*{Coordination-Heavy Case (Negative Scalability):}
Simulation time is increasing with the number of nodes, for the case of (noOfVMs = 200, noOfCloudlets = 400, isLoaded = false). This is because the cloudlets are not doing any task or no load attached to each cloudlet to perform. Hence, Hazelcast integration imposes an overhead consisting of coordination and other fixed costs, for an application for which a sequential and centralized execution is good enough. As in the success case, the time is converging here as well. Introducing further nodes will not increase the time any more, after some number of nodes. This case was optimized with the improvements discussed in the previous chapters, as the controlling case.

\paragraph*{Common Case (Positive Scalability followed by Negative Scalability):}
Simulation time is decreasing with the number of nodes steadily till a number of nodes, and then it starts to increase steadily, for the case of (noOfVMs = 100, noOfCloudlets = 175, isLoaded = true). This is one of the commonest cases, where a memory-hungry application that can hang (infinitely long time) in a single node, runs faster (10x speedup) in 2 nodes and also in 3 nodes, where further nodes may decrease the performance, due to the coordination and communication costs. In this particular example, 5 nodes was the ideal scenario and introducing the 6th node created a negative scalability. Here the communication and serialization costs start to dominate the benefits of the scalability at latter stages.

\paragraph*{Complex Case (Weird Patterns and borderline cases):}
Scenario (noOfVMs = 100, noOfCloudlets = 150, isLoaded = true) initially shows a negative scalability, followed by a positive scalability and then by a negative scalability again. Through repeating different experiments, a pattern was noticed in this rarely occurring scenario. Initially, introducing Hazelcast causes an overhead over the performance enhancements it provides, hence increasing the execution time. Then, the application starts to have the advantages of distribution and enhanced scalability, when the speedup due to distribution dominates over the initial overheads of distribution, specially the serialization and initialization costs of Hazelcast, as described in Section~\ref{sec:3perf}. Later, communication costs tend to overtake the advantages of the distribution, causing negative scalability again. These are borderline cases, where an ideal number of nodes for the distribution cannot be easily predicted.

The distinct cases of scalability can be predicted by following the performance and speed up analysis presented in Section~\ref{sec:3perf}. Among all the cases, there was a pattern, and it was possible to predict the changing scalability pattern, based on the curves for the other number of cloudlets and VMs combinations, given that the application remained unchanged. 

\subsection{Fair Matchmaking-based Cloudlet Scheduling}
A fair matchmaking-based scheduling~\cite{mm,mm2} scenario is implemented to depict a practical use case of distributed execution of simulations. While other parameters are kept constant as in the previous scenario, the number of cloudlets was changed. The workload of this execution is a matchmaking-based cloudlet scheduling. Each cloudlet and VM has a variable length or size. Each cloudlet requires the executing VM to have a minimal size, which is a function of the cloudlet length. Cloudlets search the object space to find the best fit for this specification, and bind themselves to the VM that is the best fit. While ensuring that the minimal specifications are met, cloudlets also ensure fairness, by not binding to a VM that is much larger than their specification requirements. This avoids overloading the large VMs, and schedules a fair share of cloudlets to the VMs that they are bound to, in a round robin manner.

Figure~\ref{fig:hzcloudlet} depicts the time taken for simulations with different number of cloudlets, with multiple nodes. Execution time for CloudSim was almost the same as the simulation time in a single node in $Cloud^{2}Sim$, except for very small number of objects, where the inherent overheads such as threads initializations of $Cloud^{2}Sim$ were visible. As shown by Figure~\ref{fig:hzcloudlet}, $Cloud^{2}Sim$ exhibits a positive scalability for larger simulations, handling the simulation effectively through the distributed execution. As the simulation size becomes larger with large number of VMs and cloudlets, simulation time grows exponentially due to the increasing search and matchmaking space, when running on a single instance. This exponential growth is handled and mitigated when running on multiple instances, as the execution is evenly distributed across the instances.

\begin{figure}[ht]
\begin{center}
 \resizebox{0.65\columnwidth}{!}{
  \includegraphics[width=0.65\textwidth]{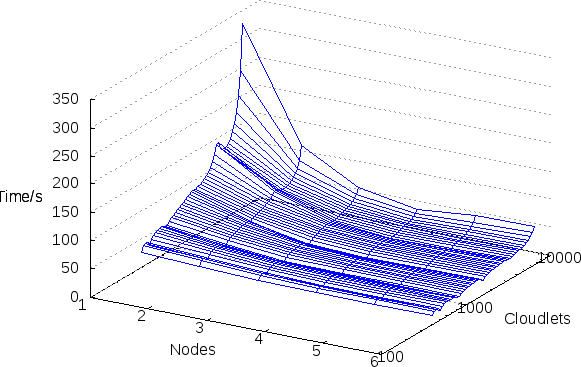}
 }
\end{center}

 \caption{Simulation Time for Matchmaking-based scheduling}
 \label{fig:hzcloudlet}
\end{figure}

Health Monitoring module of $Cloud^{2}Sim$ monitors the health of the simulator and provides the summary of the health status of the simulations during and after the simulations, according to the configurations. Maximum process CPU load monitored while the process CPU load was consistently increasing, for the above simulation scenarios is depicted by Figure~\ref{fig:cpuload}. Serializations/deserializations and communication across the cluster contribute to the higher process CPU load in simulations with multiple clusters, while the distributed execution minimizes the load due to the cloudlet scheduling simulation itself. As the CPU utilization as well as the other health parameters are monitored from the master instance, CPU utilization increases as the simulation progresses instead of blocked in synchrnonization. 

\begin{figure}[ht]
\begin{center}
 \resizebox{0.6\columnwidth}{!}{
  \includegraphics[width=0.6\textwidth]{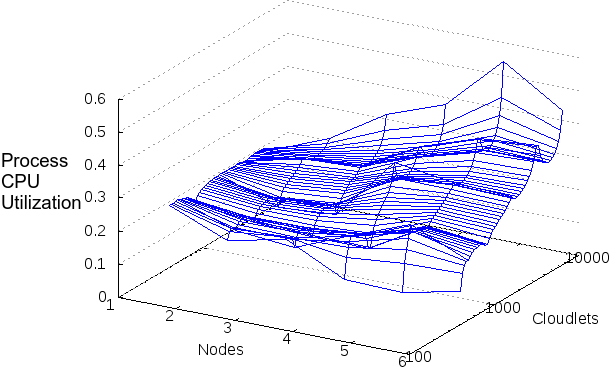}
 }
\end{center}

 \caption{Maximum Process CPU Load Monitored}
 \label{fig:cpuload}
\end{figure}

As more instances are added, simulation performs faster. The performance gain, or the percentage improvement in the simulation time for the multiple instances is shown by Figure~\ref{fig:speedupp}. $Cloud^{2}Sim$ provided a considerable performance gain to the simulations, compared to their serial execution.

\begin{figure}[ht]
\begin{center}
 \resizebox{0.6\columnwidth}{!}{
  \includegraphics[width=0.6\textwidth]{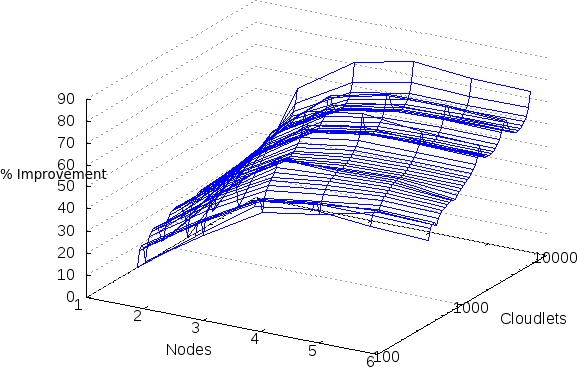}
 }
\end{center}

 \caption{Speedup - Percentage Improvement of the Distributed Execution}
 \label{fig:speedupp}
\end{figure}

Efficiency is a measure that can be used to find the ideal number of instances for a cost-effective and efficient distributed execution. Efficiency of the platform depends on the nature of the application itself. Due to the unparallelizable core simulation segments, increasing the number of instances beyond a certain number of instances will not further increase the efficiency. Figure~\ref{fig:efficiency} depicts the efficiency of the platform for the fair matchmaking-based scheduling. The graph indicates 3 or 4 as the ideal number of instances in this scenario. Efficiency was noticed to exceed 100\% in these scenarios. However, this number differs for different applications, and also is based on the scale of the applications. Larger simulations showed a higher efficiency on multiple nodes, proving that distributing smaller simulations may not be efficient.

\begin{figure}[ht]
\begin{center}
 \resizebox{0.55\columnwidth}{!}{
  \includegraphics[width=0.55\textwidth]{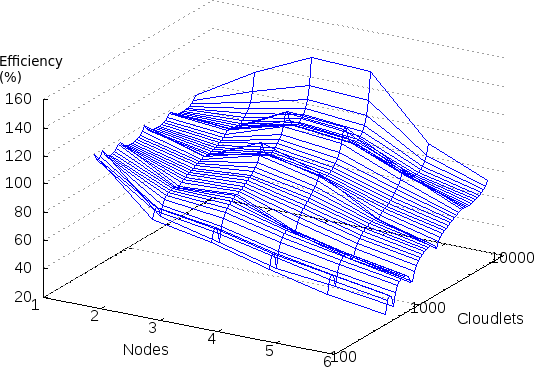}
 }
\end{center}

 \caption{Efficiency of the simulation platform with multiple instances}
 \label{fig:efficiency}
\end{figure}

\paragraph*{Hazelcast for $Cloud^{2}Sim$:}
The effectiveness of using Hazelcast to distribute the storage and execution for the simulation was evaluated by observing the overhead it imposes. Its distribution of execution and storage was measured by observing Hazelcast Management Center. Distributed objects and distributed execution were monitored by Hazelcast Management Center. 

Hazelcast Management Center indicated equal partitioning of storage across all the instances. Objects were uniformly distributed among the available Hazelcast instances, consuming almost the same amount of entry memory from each instance. Partitions of different instances were equally hit or accessed. This shows an effective partitioning of the distributed objects by Hazelcast. Figure~\ref{fig:mancenter} shows a screenshot of Hazelcast Management Center, while $Cloud^{2}Sim$ was running a sample simulation.
\begin{figure}[ht]
\begin{center}
 \resizebox{\columnwidth}{!}{
  \includegraphics[width=\textwidth]{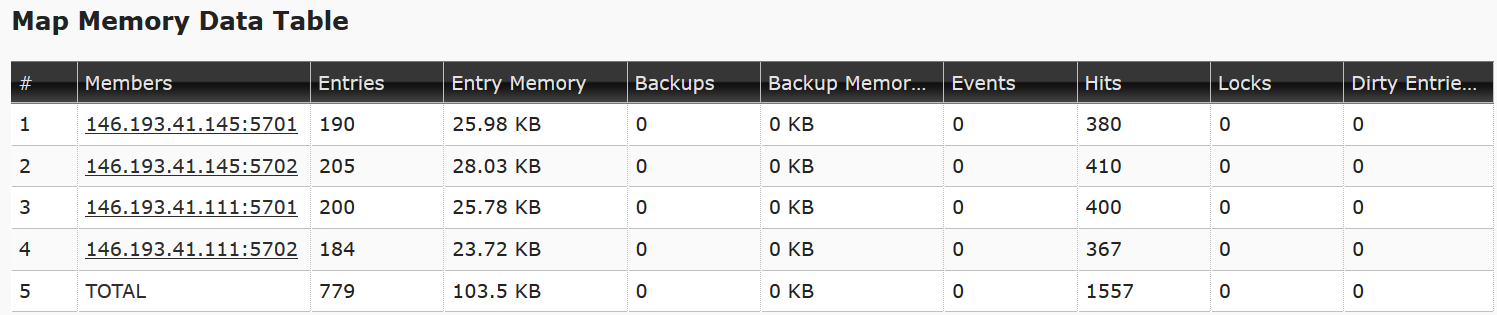}
 }
\end{center}

 \caption{Distributed Objects as Observed by Hazelcast Management Center}
 \label{fig:mancenter}
\end{figure}

\section{MapReduce Simulations}
\label{sec:5ms}
Hazelcast-based and Infinispan-based MapReduce simulator word count implementations were benchmarked against multiple big files of 6 - 8 MB, each consisting of more than 125,000 lines, having the full size up to 9.4 GB. Both implementations were observed to distribute the job uniformly across all the instances in the execution cluster. 

Figure~\ref{fig:mr} represents the time taken for both implementations on a single server with 3 map() invocations, along with the increasing number of reduce invocations with the size. Here the size is measured by the number of lines taken into consideration for the MapReduce task.

\begin{figure}[ht]
\begin{center}
 \resizebox{0.7\columnwidth}{!}{
  \includegraphics[width=0.7\textwidth]{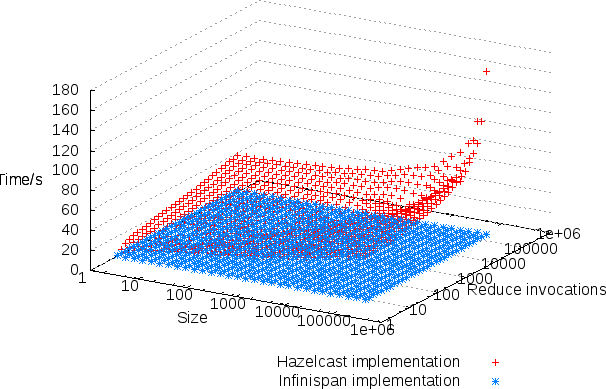}
 }
\end{center}
 \caption{Reduce invocations and time taken for different sizes of MapReduce tasks}
 \label{fig:mr}
\end{figure}

The results showed Infinispan outperforming Hazelcast by 10 to 100 folds. As verbose mode monitors and logs each of the execution steps and configurations, executions were slower in verbose mode, where the execution in non-verbose mode was much faster. Infinispan based simulator was still fast, even when operating verbose. Infinispan MapReduce implementation is matured. Hazelcast MapReduce implementation is young, and still could be inefficient. Infinispan performs well in a single-node mode, as it operates better as a local cache. Hazelcast is optimized for larger set ups with very high number of real server nodes, and probably Hazelcast could outperform Infinispan, when larger number of nodes (such as 50) are involved. MapReduce executions are easily parallel and distributed by nature, even in a single node. This is not the case for general applications like CloudSim simulations. Infinispan model is perfect for a single node (or even a few node) MapReduce tasks.

\subsection{Infinispan MapReduce Implementation}
Infinispan implementation was tested for its scalability, with the same MapReduce job distributed to different number of nodes. Figure~\ref{fig:infmap} shows the scaling of Infinispan MapReduce implementation to multiple nodes, with the time taken to execute different number of map() invocations. Number of reduce() invocations was kept constant at 159,069. Number of map() invocations is equal to the number of files present in the word count execution used. Hence, the number of files were increased for different scenarios. As the number of instances were increased, the jobs were distributed to the available instances.

\begin{figure}[ht]
\begin{center}
 \resizebox{0.7\columnwidth}{!}{
  \includegraphics[width=0.7\textwidth]{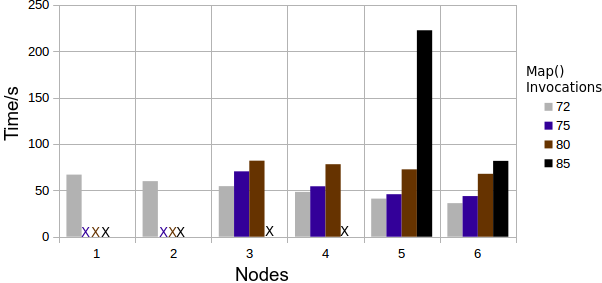}
 }
\end{center}
 \caption{Distributing the Infinispan MapReduce execution to multiple nodes}
 \label{fig:infmap}
\end{figure}

When the number of map() invocations was increased, jobs started to fail in single instance, due to the out of memory (java.lang.OutOfMemoryError: Java heap space) issue. Further, garbage collection (GC) overhead limit was exceeded in some scenarios. These issues prevented larger invocations to execute in smaller number of instances. When the number of instances was increased, the jobs that failed started to execute successfully and a positive scalability was observed. These evaluations prove that memory and processing requirements increase as the number of map() and reduce() invocations are increased. Further, distributing the execution enables larger executions, and makes the executions faster.

\subsection{Hazelcast MapReduce Implementation}
As Hazelcast based MapReduce simulator was slow when run in a single mode, it is tested on 1 - 6 nodes in verbose mode to check the improvements in execution time. One node starts the MapReduce simulator, where other nodes start the $Initiator$ class, which just connects to the cluster and executes the logic fractions sent by the master. 

All the Initiator nodes were started to form a cluster, before starting the instance running the simulator. Time taken for different sizes of the task to run on different number of instances is shown by Figure ~\ref{fig:distrhzmr}. Number of map() invocations was kept constant at 3, while increasing the number of reduce() invocations. Infinispan with single node was noticed to be still faster than all 6 nodes running MapReduce in Hazelcast. 

\begin{figure}[ht]
\begin{center}
 \resizebox{0.6\columnwidth}{!}{
  \includegraphics[width=0.6\textwidth]{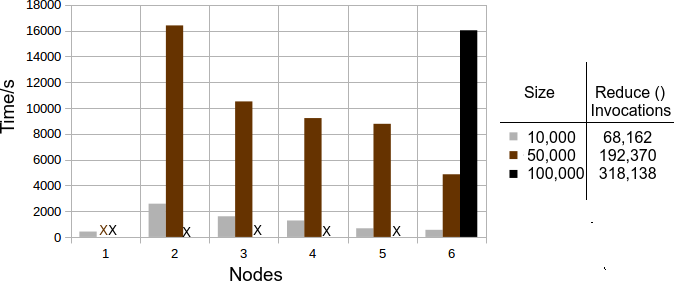}
 }
\end{center}
 \caption{Distributing the Hazelcast MapReduce execution to multiple nodes}
 \label{fig:distrhzmr}
\end{figure}

For the size of 10,000 (68,162 reduce() invocations), Hazelcast running on a single instance was fast enough, and distributing the execution to multiple nodes started with a considerable negative scalability. This is because the communication and coordination costs were higher than the improvements from the distributions. However, positive scalability, though not significant, was achieved when more than 8 instances were used, as shown by Table~\ref{table:hz12}. Up to 2 Hazelcast instances were executed from each of the nodes during this. This shows that even for smaller applications, distribution may be advantageous overtaking the communication and other costs introduced by distributing the execution.

\begin{table}[!t]
\caption{Time (sec) taken for multiple Hazelcast instances to execute the same task}
\label{table:hz12}
\begin{tabular}{|c||c| |c| |c||c| |c| |c||c| |c| |c|}
\hline
\textbf{Number of Hazelcast Instances} & \textbf{1} & \textbf{2} & \textbf{3} & \textbf{4} \\
\hline
\textbf{Time taken (sec)} & 416.687 & 2580.087 & 1600.655 & 1275.664 \\
\hline
\hline
 & \textbf{6} & \textbf{8} & \textbf{10} & \textbf{12}\\
\hline
 & 553.296 & 432.926 & 320.055 & 312.414 \\
\hline 
\end{tabular}
\end{table}

The sample application failed to run on single node for the size of 50,000 (192,370 reduce() invocations) due to the heap space limitations. It ran smoothly on 2 instances, and showed a perfect positive scalability, when the nodes were joined to the cluster up to 6. The application failed to run on a single node for the size of 100,000 (318,138 reduce() invocations), due to the out of memory issue in heap space. The issue persists even when the cluster size was increased up to 5 nodes. The application ran successfully only when 6 nodes were involved. The last two cases clearly show the requirement of distributed MapReduce simulations for larger tasks, as a single or a fewer nodes in the cluster were proven to be insufficient for the higher memory requirements of the MapReduce tasks.

\paragraph*{Bugs and Limitations:}
A few critical bugs were encountered during the evaluations of Hazelcast MapReduce implementation. If a new Hazelcast instance joins a cluster that is running a MapReduce job, it is noticed to crash the instance running the MapReduce task and hence failing the MapReduce task\footnote{\url{https://github.com/hazelcast/hazelcast/issues/2354}}. This was caused by the newly joined instance not knowing the supervisor of the job, due to a missing null-check, according to the core Hazelcast/MapReduce developer. As a work-around, the master instance that starts the MapReduce jobs was started and joined the cluster, only after all the Initiator instances have started and formed the cluster. This prevented incorporation of the Hazelcast-based auto scaling and adaptive scaling that were already implemented during this project. Moreover, in a long running heavy task, instances were noticed to leave the cluster, to exhibit a split-brain syndrome\footnote{\url{https://github.com/hazelcast/hazelcast/issues/2359}}. This limited the usability of the MapReduce implementation to shorter MapReduce jobs. These issues were reported to the Hazelcast issue tracker. Eventually, the reported bugs have been addressed in the later versions of Hazelcast. 

\section{$Cloud^{2}Sim$ and Related Work}
\label{sec:5rw}
Here, we will compare the features of $Cloud^{2}Sim$ with the other related work.

\paragraph*{$Cloud^{2}Sim$ and CloudSimEx:}
$Cloud^{2}Sim$ and CloudSimEx are two simulators or simulator extensions that are built on top of CloudSim to enhance the abilities of CloudSim simulator. While $Cloud^{2}Sim$ focuses more on the scalability and performance of CloudSim to perform larger simulations, CloudSimEx focuses on providing extension points to CloudSim, adding a new set of features. While there are some overlaps, the features provided by these two products are complimentary as listed by Table~\ref{table:sim2ex}, and CloudSimEx extensions can basically be used with $Cloud^{2}Sim$.
\begin{table}[!t]
\caption{CloudSimEx and $Cloud^{2}Sim$ Features Comparison}
\label{table:sim2ex}
\begin{tabular}{|c||c| |c|}
\hline
& \textbf{CloudSimEx} & \textbf{$Cloud^{2}Sim$}\\
\hline
\hline
\textbf{Simulation Scenarios} & & \\
\hline
MapReduce Simulation & \checkmark & \checkmark \\
Billing for cloud IaaS (AWS, Google, ..) & \checkmark & X \\
Modelling disk operations & \checkmark & X \\
Web session modeling & \checkmark & X \\
Modeling network latencies/delays & \checkmark & X \\
Modeling auto scaling & \checkmark & X \\
\hline
\hline
\textbf{Feature Enhancements} & & \\
\hline
Running multiple experiments in parallel & \checkmark & \checkmark \\
 & (different processes) & (different clusters) \\
Better logging (to CSV, ..) & \checkmark & X \\
Geographical location of servers & \checkmark & X \\
 & (using GeoLite2) &  \\
Configure using configuration files & X & \checkmark \\
Optional loads for cloudlets to perform & X & \checkmark \\
Health monitoring & X & \checkmark \\
\hline
\hline
\textbf{Scalability Improvements} & & \\
\hline
Improved concurrency & X & \checkmark \\
Distributed execution & X & \checkmark \\
Auto scaling of the simulations & X & \checkmark \\
Scaling the simulations on an IaaS & X & \checkmark \\
& & (on aws) \\
Adaptive Scaling / Cycle Sharing & X & \checkmark \\
\hline
\end{tabular}
\end{table}

\paragraph*{$Cloud^{2}Sim$ and iCanCloud:}
$Cloud^{2}Sim$ and iCanCloud are two simulators that focus on large cloud simulations. iCanCloud finds Java language as a limiting factor on simulations higher than 2 GB on 32 bit platforms, and hence is built on C++. $Cloud^{2}Sim$ tackles the memory and processing limitations by exploiting the in-memory data grids over multiple nodes. The features and characteristics of these simulators are compared  by Table~\ref{table:ican}.
\begin{table}[!t]
\caption{iCanCloud and $Cloud^{2}Sim$ Features Comparison}
\label{table:ican}
\begin{tabular}{|c||c| |c|}
\hline
& \textbf{iCanCloud} & \textbf{$Cloud^{2}Sim$}\\
\hline
\hline
\textbf{Development} & &\\ 
\hline
Core Component / & OMNeT++ & CloudSim \\
Simulation Library & Discrete Event Simulator & Cloud Simulator \\
\hline
Strategy for & Using C++ & Using Java In-Memory Data Grids \\
Huge Simulations & & (Hazelcast and Infinispan) \\
\hline
Distributed Execution & MPI  & Hazelcast and Infinispan \\
on Multiple Nodes & (Work-in-progress) & (Distributed Cache)\\ 
\hline
\hline
\textbf{Scalability} & &\\ 
\hline
Vertical Scalability & High & High \\
& (By design) & (Concurrency) \\
\hline
Horizontal Scalability & Work-in-progress & High \\
& (MPI) & (Hazelcast and Infinispan) \\
\hline
Elasticity & X & \checkmark \\
\hline
\end{tabular}
\end{table}

\paragraph*{Summary:}
$Cloud^{2}Sim$ shows a considerable performance improvement for large simulations, compared to CloudSim. It was possible to execute the memory-hungry applications that would not run in a single node, as the required memory to store the objects exceeded the available memory in any single node in the cluster. Cloud and MapReduce simulations that have large resource requirements such as memory and CPU, execute faster and more effectively, over multiple instances. Infinispan based MapReduce implementation is proven to be much faster than the Hazelcast based implementation, based on the samples that are tried.
\chapter{Conclusion and Future Work}
\label{chap:cf}
\section{Conclusion}
\label{sec:6c}

Simulation tools try to portray a geo-distributed decentralized environment using network and topology simulation code that is serial and manipulating a large global state that is considered consistent. Hence, their performance is far from ideal. Typically, cloud and MapReduce simulators are sequential, and thus run on a single computer, where computer clusters and in-memory data grids can be leveraged to execute larger simulations that cannot be executed on a single computer. Even the simulations that can run on a single node can take advantage of more resources from the cluster, that it can run faster and more effectively. The cycle sharing model can be utilized to provide means of sharing the resources across the simulation instances, allowing multiple independent simulations to execute in parallel, in a multi-tenanted manner.

A scalable middleware platform for concurrent and distributed cloud and MapReduce simulations can leverage an existing cloud simulator, whilst exploiting the in-memory data grid platforms for an elastic environment, deploying an adaptive scaling strategy inspired by the volunteer computing model. $Cloud^{2}Sim$ presents an architecture that enables the execution of larger simulations in a cluster, that cannot be run on single nodes due to the requirement of huge heap space, and the hindrance of long execution times. $Cloud^{2}Sim$ scales reasonably well, and distributes the storage and execution almost uniformly among all the instances. $Cloud^{2}Sim$ has the advantages and wide applicability of CloudSim while being efficient, faster, customizable, and scalable. By virtue of being elastic and adaptive, it is also itself cloud-ready, and can be the basis of a truly concurrent and distributed Simulation-as-a-Service for Cloud and MapReduce simulations.
  %
  %

\section{Future Work}
\label{sec:6fw}
While designing and developing $Cloud^{2}Sim$, further extension points were noticed. They can be implemented as part of $Cloud^{2}Sim$, or as independent entities, as discussed below.

\paragraph*{State-aware Adaptive Scaling:}
State is shared among the instances using distributed atomic objects in the adaptive scaler. Currently this is only used for a very small state that drives the scaling decisions. This also can be extended to share the state of the simulation or application, such that a new instance could join the execution at a later time for a specific sub-task based on the load, and could be terminated upon the completion of the sub-task, or when the load goes down. State-awareness may increase the resource utilization and efficiency of the simulator, with possibilities of predicting higher loads based on the state of the simulations.

\paragraph*{Lazy Loading:}
Serialization and deserialization impose a considerable overhead on the simulator, due to the extensive use of complex objects in the simulations. A lighter version of these objects can be used along with a lazy loading pattern~\cite{veiga2002incremental}, where the complete objects will be initialized and loaded only when one of their properties is accessed. Lighter versions of the actual CloudSim data structures will function as virtual proxies of the objects. This will mitigate the serialization costs. Moreover, lighter objects with fewer parameters and properties could replace the CloudSim objects, for a few simulations.

\paragraph*{Infinispan based Cloud Simulations:}
Infinispan is integrated into the compatibility layer such that it can be used to implement an extension to CloudSim, following the current implementation of $Cloud^{2}Sim$ Cloud Simulations. Currently, only a Hazelcast based cloud simulator is available, though further research showed that the design of Hazelcast based cloud simulations would fit the simulations based on Infinispan as well. Following the same design, a complete distributed cloud simulator with any other in-memory data grids can also be built, as loose coupling is maintained between the data grid libraries and $Cloud^{2}Sim$.

\paragraph*{Generic Elastic Middleware Platform:}
MapReduce implementations stand as an extension and proof that the same distributed execution model can be extended beyond cloud simulations. The design of the elastic middleware platform can be extended to other cloud and MapReduce simulators, as the design of Hazelcast and CloudSim based $Cloud^{2}Sim$ distributed cloud simulator is not tightly coupled to CloudSim or Hazelcast. Moreover, the adaptive scaler design suits many other applications (such as MapReduce executions, MPI workflows, and scientific workflows), not just limited to simulations. Hence this can be extended to be used on any application that has an elastic scaling requirement.


  %
  %
\bibliographystyle{chicago}
\bibliography{document}




  %
  %

\appendix

  %
  %

%
%

  %
  %

\chapter{$Cloud^{2}Sim$ Configurations}
\label{ch:commodoD}

$Cloud^{2}Sim$ is configured for multiple simulations using the configuration file $cloud2sim.properties$, present in the conf folder. The properties file provides a unified configuration for all the use cases of $Cloud^{2}Sim$. Relevant sections of the properties file are uncommented based on the use case, where the others are removed or left commented out.

\paragraph*{Infinispan Configuration:}
Infinispan configuration is managed by $infinispan.xml$. For the distributed operation, transport is configured using either TCP or UDP. The configurations are handled by pointing to the respective JGroups configuration files, $jgroups-tcp-config.xml$ and $jgroups-udp-config.xml$.

\paragraph*{Hazelcast Configuration:}
Hazelcast properties section in the configuration file provides the location of the hazelcast.xml file, which is available in two sample versions, configured for cloud simulations and MapReduce simulations. Name of the simulation cluster and the supportive/sub cluster are also configured using the configurations. Parallel experiments can be run from the same computer or network of computers, by changing the value of the mainCluster property. In a multi-tenanted environment, multiple clusters can be started by the $Coordinator$, and all the respective cluster names should be indicated.

Parallel supporter clusters can also co-exist, by changing the value of the subCluster property. However, these values can also be changed programmatically by the running simulator instances. Though instances cannot change their cluster group after they have started, instances can be joined to a cluster specified programmatically or by the configuration, at the start up time.
 
{\fontsize{10}{10}\selectfont
\begin{lstlisting}
# For Map-Reduce Simulations
hazelcastXml=conf/mapreduce/hazelcast.xml
## For Cloud Simulations
#hazelcastXml=conf/hazelcast.xml

# Name of the simulation cluster. Default is "main". Change this for parallel simulations.
# Uncomment and use different cluster names for multi-tenanted deployments.
mainCluster="main"
# mainCluster2="main2"
# mainCluster3="main3"

# Name of the supportive cluster. Default is "sub".
# The cluster used for scaling and health decisions.
# Example use case: IntelligentAdaptiveScaler.
subCluster="sub"

\end{lstlisting}
}

\paragraph*{MapReduce Configuration:}
Since MapReduce simulator has two implementations, Hazelcast-based and Infinispan-based, both implementations could be configured using MapReduce simulation properties section in the $cloud2sim.properties$. MapReduce size, the location of the folder that contains the load files, and verbosity level are configured. MapReduce simulation could be configured with health monitoring and scaling. Certain detailed logs such as a detailed progress update on the MapReduce simulations are shown by the MapReduce simulator only when the isVerbose property is set to true in $cloud2sim.properties$. However, verbose mode should be used only when it is required as it takes more resources and slows down the execution. 

The property $filesRead$ can be used to limit the number of files read from the folder, with the default value 0 reads all the files in the directory. The default load folder for MapReduce jobs, $conf/mapreduce/load$ contains 3 text files of 6 - 8 MB size with more than 128 460 lines. Users may replace these files or add more or bigger files. They may also use another folder as the load folder.
 
{\fontsize{10}{10}\selectfont
\begin{lstlisting}
# up to 128 460, by using the 3 default files in conf/mapreduce/load
mapReduceSize=100000

# Verbosity level
isVerbose=true

# Load for the MapReduce jobs
loadFolder=conf/mapreduce/load

# Proportional to the invocations of map().
filesRead=3
\end{lstlisting}
}

\paragraph*{Cloud Simulation Configuration:}
This is used to configure the sample cloud simulation applications. Cloud simulations require the $noOfUsers$ property defined, to be able to start the execution. At least one cloud user should be present to enable the cloud simulations. Hence, if the property is not set, cloud simulation does not execute. Multiple instances can be started at once from the same node by specifying a value for the $simultaneousInstances$ property. $noOfExecutions$ property makes the simulation wait till the specified number of instances join the cluster.

{\fontsize{10}{10}\selectfont
\begin{lstlisting}
noOfUsers=200
noOfDatacenters=15
noOfHosts=20
noOfVms=200
noOfCloudlets=400

# How many Hazelcast instances to be spawn at once.
simultaneousInstances=1

# How many executions are run in parallel.
# Simulation waits till the specified number of instances have joined the simulation cluster.
noOfExecutions=2

# Enter workload to load the cloudlets with a series of heavy mathematical operations.
withWorkload=false
\end{lstlisting}
}

\paragraph*{Health Monitor Configuration:}
Health checking is disabled by default. It can be enabled by including the $timeBetweenHealthChecks$ property.

{\fontsize{10}{10}\selectfont
\begin{lstlisting}
# in seconds
timeBetweenHealthChecks=10

# Params: Means of checking the health
highThresholdProcessCpuLoad=0.15
lowThresholdProcessCpuLoad=0.02
\end{lstlisting}
}

\paragraph*{Dynamic Scaling Configuration:}
Cloud and MapReduce simulations can be configured with the health monitoring, where dynamic scaling can be configured for Hazelcast based MapReduce and cloud simulations. Properties such as maximum instances that are spawned or allowed in the cluster, time between scaling, and the mode of scaling are specified.

{\fontsize{10}{10}\selectfont
\begin{lstlisting}
maxNumberOfInstancesToBeSpawned=3

timeBetweenScalingDecisions=60

## mode: how to scale based on the load.
# auto - for scaling out by spawning new instances inside the same computer or in AWS EC2.
# adaptive - for scaling out by adding instances adaptively into the simulation cluster.
scalingMode=adaptive

\end{lstlisting}
}

\end{changemargin}

  %
  %
\begin{singlespace}

\def\indexname{Index}             
\printindex\cleardoublepage

\end{singlespace}

\end{document}